\documentclass[jmp]{revtex4-1}
\usepackage{amsmath}
\usepackage{amsfonts}
\usepackage{graphicx}

\newcommand{\tx}[1]{\text{#1}}

  \def\rfr{{\mathcal{R}}}
  \def\bigo{{\textrm{O}}}
  \def\sz{{\mbox{\tiny $0$}}}

  \def\sh{{\mbox{\small $\frac{1}{2}$}}}

  \def\rr{{ {\bf r} }}
  \def\ro{{ {\bf r}_{\sz} }}
  \def\rmn{{ \rr_{m,n} }}
  \def\ron{{ \rr_{ \sz,n } }}
  \def\ex{{ {\bf e}_{1} }}
  \def\ey{{ {\bf e}_{2} }}
  \def\ez{{ {\bf e}_{3} }}
  \def\kk{{ {\bf k} }}
  \def\hz{{ H_{\sz} }}
  
  \def\kx{{ k_{x} }}
  \def\kz{{ k_{z} }}
  \def\kzn{{k_{z,-1}}}
  \def\qzn{{q_{z,-1}}}
  \def\kzm{{  k_{z,m} }}
  \def\qzm{{  q_{z,m} }}
  \def\uu{{ {\bf u} }}
  \def\ee{{\varepsilon}}
  \def\eec{{\varepsilon_{c}}}

  \def\eg{\mathcal{E}}

  \def\eik{{ e^{ i {\bf k} \cdot \rr } }}

  \def\eo{{ E_{\omega}  }}

  \def\cpm{{ \eo (  \pm  h\ez ) }}

  \def\cph{{ \eo (   - h\ez ) }}
  \def\cpa{{ \eo (   a \ex + h\ez ) }}

  \def\pst{{\Phi_{\sz}}}
  \def\pap{{\Phi^{+}}}
  \def\pam{{\Phi^{-}}}
  \def\pamp{{\Phi^{\pm}}}
  \def\phc{{\Phi_{c}}}
  \def\phs{{\Phi_{s}}}

  \def\php{{\Phi_{\ast}}}
  \def\dlo{{\delta_{\sz}}}
  \def\dlot{{\delta^{2}_{\sz}}}

  \def\pn{{P_{2n-1}}}
  \def\ccp{{\mathcal{C}^{+}}}
  \def\ccc{{\mathcal{C}_{c}}}

\begin{document}
\title{Electromagnetic bound states in
the radiation continuum
for periodic double arrays of subwavelength
dielectric cylinders}
\author{ R\'{e}my F. Ndangali and Sergei V. Shabanov \\ Department of Mathematics, University of Florida, Gainesville, FL 32611, USA}

\begin{abstract}
Electromagnetic bound states in the radiation continuum are studied for periodic double arrays of subwavelength dielectric cylinders in TM polarization. They are similar to localized waveguide mode solutions of Maxwell's equations for metal cavities or defects of photonic crystals, but, in contrast to the latter, their spectrum lies in the radiation continuum. The phenomenon is identical to the existence of bound sates in the radiation continuum in quantum mechanics,
discovered by von Neumann and Wigner. In the formal
scattering theory, these states appear
 as resonances with the vanishing width. For the system studied, the bound states are shown to exist at specific distances between the arrays in the spectral region where one
or  two
diffraction channels are open. Analytic solutions are obtained for all bound states (below the radiation
continuum and in it) in the limit of thin cylinders (the cylinder radius is much smaller than the wavelength).
The existence of bound states is also established
in the spectral region where three and more diffraction
channels are open, provided the dielectric constant and radius of the cylinders are fine-tuned.
The near field and scattering resonances of the structure are investigated when the distance between the arrays varies in a neighborhood of its critical values at which the bound states are formed. In particular, it is shown that the near field in the scattering process becomes significantly amplified in specific regions of the array as the distance approaches its critical values. The effect may be used to control optical non-linear effects by varying the distance between the arrays near its critical values.
\end{abstract}

\maketitle

\section{Introduction}
\label{sec:0}
Scattering problems in quantum mechanics and Maxwell's theory exhibit a great deal of similarities. The dielectric constant of an electromagnetic scattering structure is analogous to a scattering potential in the Schr\"odinger equation. If the potential is real and bounded from above, then solutions of the wave equation (Schr\"odinger's or Maxwell's) appear in two kinds \cite{b2}. There are radiation modes that carry the energy flux to the spatial infinity and modes trapped in some (bounded) spatial region. The former are known as scattering states. They do not have a finite $L_2$ norm, and their spectrum is continuous (the radiation continuum). In contrast, solutions of the second kind have a finite $L_2$ norm and a discrete spectrum which typically lies below the radiation continuum. In quantum mechanics, these are bound states. In Maxwell's theory, they are electromagnetic waves localized, e.g., in metal cavities or in defects of photonic crystals. It was first proved by von Neumann and Wigner \cite{b1} in 1929 that, in quantum mechanics, there may exist bound states in the radiation continuum (i.e., there are localized solutions of the Schr\"odinger equation whose energies lie in the continuum part of the system spectrum). An apparent physical peculiarity of such states is that, despite the boundedness of the potential barrier, they do not decay into the scattering modes through the tunneling, which is rather counterintuitive
and unusual. An example constructed by von Neumann and Wigner was somewhat artificial from the physical point of view, and no significant applications of this phenomenon immediately followed their discovery. However, much later unusually long-lived resonances were observed in atomic physics \cite{b3} and the question about a physical mechanism  of their stability arose.\\
\indent From the mathematical point of view, scattering resonances can be associated with solutions of the wave equation that satisfy the outgoing wave boundary condition at the spatial infinity, i.e., in the asymptotic region, the solutions look like waves outgoing from the scattering structure \cite{b2}. Clearly, such solutions cannot preserve energy (or probability) flux and, hence, do not exist for real frequencies (or energies). So they have complex eigen-frequencies and decay exponentially with time. Due to linearity of the wave equation, a solution of a scattering problem is a linear superposition of the incident wave and scattered waves that satisfy the outgoing wave boundary condition. The resonant solutions may therefore be used to construct the scattered wave, and the flux conservation is achieved by balancing the incoming flux of the incident wave and the outgoing flux carried by scattered waves. It was proved then that the resonant scattering amplitude may be approximated by the celebrated Breit-Wigner formula \cite{b4,b2} $A(\omega) \sim i\Gamma/(\omega -\omega_r + i\Gamma)$ where $\omega$ is the incident wave frequency, and $\omega_r -i\Gamma$ is the eigen-frequency of the resonant mode. If the imaginary part $\Gamma$ is small ($\omega_r \gg \Gamma$), then the scattering cross-section $\sigma (\omega)\sim |A(\omega)|^2$ exhibits a sharp maximum (of width $\Gamma$)
at frequencies
near  $\omega_r$. Thus, resonances with small $\Gamma$ can be associated with quasi-bound states, excitations with a finite lifetime $\sim 1/\Gamma$, that decay by emitting a nearly monochromatic radiation of frequency $\omega_r$. In this picture, the bound states in the radiation continuum may be understood as scattering resonances with the vanishing width. If a system allows for two close resonances whose positions and widths depend continuously on a physical parameter of the system that can be varied, then there might exist critical values of this parameter at which one of the resonances turns into a bound states \cite{b5,b6,b7}. This is the subject of the theory of coupled resonances developed in quantum scattering theory \cite{b4}. The conditions under which the width of one of the coupled resonances vanishes were established. The theory of coupled resonances provided an explanation of the existence of the aforementioned unusually long-lived resonances observed in atomic systems.\\
\indent Resonant scattering properties of gratings and periodically structured thin films were observed long time ago in optics \cite{b8,b9,b10}. Typically, a grating would exhibit sharp resonances near its diffraction thresholds \cite{b14}. However an explanation based on quasi-bound states came much later. Naturally, the Breit-Wigner theory can be extended to the electromagnetic scattering, and coupled resonances might be expected to exist in photonic structures \cite{b12,b17}. A simple example is provided by two parallel identical gratings. Each grating supports a  quasi-stationary electromagnetic mode localized in its vicinity (a trapped mode), which can be excited by an incident radiation. In the zero order diffraction, this mode has only one decay channel. Therefore one might anticipate a formation of a wave guiding mode confined between the structures at a certain distance between the gratings. If the trapped mode associated with each grating decays by emitting a monochromatic radiation outgoing both ways along the normal to the structure, then there might exist a distance at which the waves outgoing from each structure interfere destructively in the asymptotic region, thus producing a localized stable solution (the energy flux outgoing from the structure vanishes). This argument can be made rigorous by considering the Fabri-Perot interferometer whose interfaces have reflection and transmission coefficients described by the Breit-Wigner theory \cite{b12,b17}. When the distance gets smaller, the interference effects of the near field of the trapped modes becomes significant, and the Fabri-Perot argument becomes inapplicable as it applies only to the modes carrying the energy flux, i.e., radiation modes. Maxwell's equations must then be studied. Bound states in the radiation continuum were indeed found by numerical studies of the system of two gratings  in the spectral region where only one diffraction channel is open. They are formed at a discrete set of distances that follows the pattern predicted by the aforementioned Fabri-Perot argument for large distances and substantially deviates from it when the distance is of order the grating period \cite{b12}. Also, localized waveguide solutions were argued to exist in the same spectral region by studying Maxwell's equations for layers whose dielectric function is periodic in one direction along the layer and translational invariant along the other one \cite{b18}.\\
\indent Here bound states in the radiation continuum are studied analytically for a system of two arrays of parallel dielectric cylinders. The approach is based on the resonant scattering theory \cite{b2,b13} where the bound states are identified as resonances with the vanishing width (the distance between the arrays is a physical parameter which regulates the coupling of the resonances). The study is carried out for the whole radiation spectrum range. In addition to the bound states in the zero-order diffraction (which were known to exist from the early numerical studies), bounds states are found in the spectral range where two diffraction channels are open. Analytic solutions of Maxwell's equations for all the bound states are given in the limit when the cylinder radius is much smaller than the period of the structure. The system is shown
to have bound states below the radiation continuum whose
explicit form is also found.\\
\indent The interest to subwavelength periodic structures has been revitalized due to a recent technological progress in manufacturing such structures for the optical frequency range \cite{b14}. An ultimate goal is to develop a technology for the so-called all optical data processing \cite{b15,b16}. One way or the other, this task amounts to finding physical mechanisms to control optical non-linear effects in photonic structures as, e.g., a typical logical element in data processing requires a device similar to a transistor. As a potential application, here the near field of  resonant excitations is studied as a function of the distance between the arrays that varies in the vicinity of its critical values at which the bound states are formed (the width of one of the resonances vanishes). It is shown that the near field can be significantly amplified (as compared with the incident radiation field) in some particular regions of the structure as the distance approaches its critical values. The effect can therefore be used to enhance optical non-linear effects in a controllable way.\\
\indent The paper is organized as follows: a classification of solutions of Maxwell's equations that is convenient for identifying the bound states in the radiation continuum in the system studied is described in Section~\ref{sec:1}. Explicit solutions for the bound states below the radiation continuum as well as in the radiation continuum when one diffraction channel is open are given in Section~\ref{sec:2}. The effect of the near field amplification for a resonant scattering when the distance between the arrays  is close to one of its critical values at which the bound states are formed is investigated in Section~\ref{sec:7}. Lastly, the existence of bound states when more than one diffraction channel are open is established in Section~\ref{sec:5}. Whenever possible, the technical details are moved to the appendix.

\section{Scattering theory and classification of the fields}
\label{sec:1}
\begin{figure}[h t]
    \centering
    \includegraphics{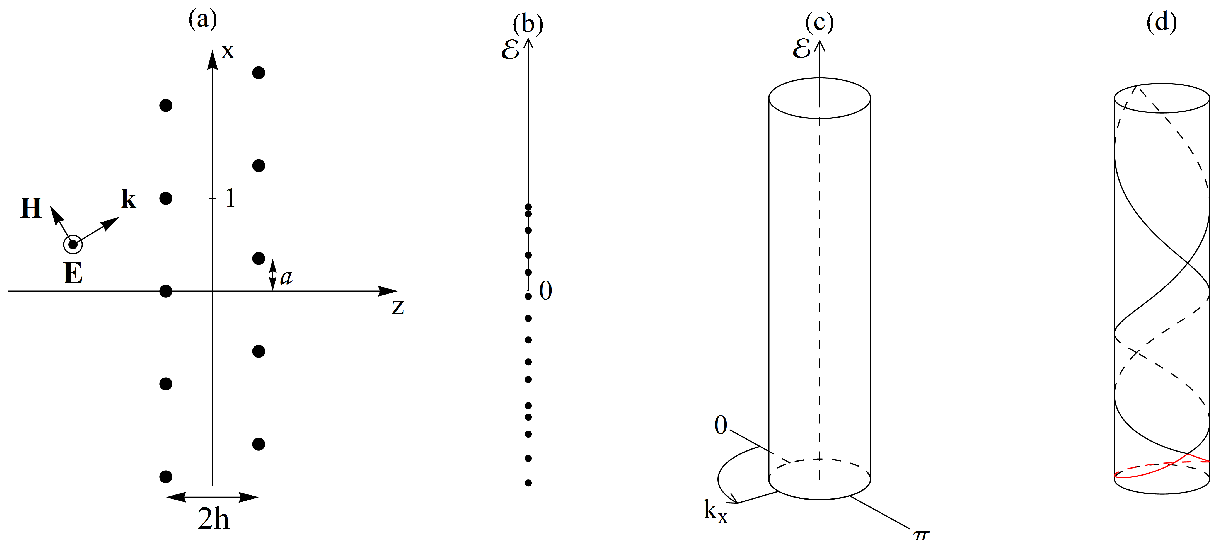}
    \caption{Panel (a): Double array of dielectric cylinders. The unit of length is the array period. The axis of each cylinder is parallel to the y-axis and is at a distance $h$ from the x-axis.\newline
    Panel (b): The energy spectrum for a Schr\"{o}dinger equation with radially symmetric potential consists of a discrete spectrum of bound states with negative energies and the radiation continuum. The latter may contain additional bound states; these are the bound states in the radiation continuum.\newline Panels (c) and (d): Harmonically time dependent solutions to Eq.(\ref{eq:1}) are labeled by points of the spectral cylinder as explained in text. The diffraction thresholds $\eg_{\pm n}$ are rings that partition the cylinder into sections corresponding to a fixed number of open channels. In particular, $\eg_{0}$ is the threshold for the radiation continuum below which there can be no scattering states.}
    \label{fig:1}
    \end{figure}

The system considered is sketched in Fig.~\ref{fig:1}(a). It consists of an infinite double array of parallel, periodically positioned, dielectric cylinders suspended in the vacuum \cite{b12}.  The cylinders are assumed to be non-dispersive with a dielectric constant $\eec>1$. The coordinate system is set so that the cylinders are parallel to the $y-$axis, the structure is periodic along the $x-$axis, and the $z-$axis is normal to the structure. The unit of length is taken to be the array period and the arrays have a relative mismatch $a\in [0,\sh]$ along the x-axis.
    \\
\indent The structure is illuminated by a linearly polarized monochromatic beam with the electric field parallel to the cylinders (TM polarization). In the given settings, Maxwell's equations are reduced to the scalar wave equation for a single component of the electric field, denoted $E$, in the $x,z-$plane:
  \begin{equation}
  \frac{\ee}{c^2}\partial_{t}^2 E-\triangle E=0\,,
  \label{eq:1}
  \end{equation}
where the dielectric function $\ee$ is piecewise constant; it is equal to $\eec>1$ on the scatterers and to $1$ otherwise. For a harmonic time dependent electric field $E(\rr,t)=\eo(\rr) e^{-i \omega t}$, the amplitude $\eo(\rr)$ satisfies the equation
  \begin{equation}
  H \eo\equiv[-\triangle+k^2 (1-\ee)] \eo=k^2 \eo\,,
  \label{eq:2}
  \end{equation}
where $k^{2}={\omega^2}/{c^2}$ is the spectral parameter. This equation is similar to the Schr\"{o}dinger equation with an attractive (negative) potential $1-\ee$. If $\kk$ is the wave vector of the incident radiation, then the periodicity of the structure requires that  solutions to Eq.(\ref{eq:2}) satisfy Bloch's theorem,
  \begin{equation}
  \eo(\rr+\ex)=e^{i k_{x} } \eo(\rr)\,,
  \label{eq:3}
  \end{equation}
where $\mathbf{k}=k_{x} \ex+ k_{z} \ez$ and $\{ \ex,\ey,\ez \}$ are the unit vectors along the coordinate axes. Consequently, the wave vectors of the scattered radiation can only have the following form
  \begin{equation}
  {\bf k}^{\pm}_{m}=(k_{x}+2\pi m) \ex\pm \sqrt{k^2-(k_{x}+2\pi m)^2}\ez\,, \qquad m=...-2,-1,0,1,2...
  \label{eq:4}
  \end{equation}
with the convention $\sqrt{k^{2}-(\kx+2\pi m)^{2}}=i\sqrt{(\kx+2\pi m)^2-k^2}$ if $k^{2}<(\kx+2\pi m)^{2}$. The values of $m$ for which
$k^{2}\geq (\kx+2\pi m)^{2}$ label {\it open diffraction channels}, all other values of $m$ correspond
to {\it closed}  channels. If, for instance, the incident radiation
propagates in the positive $z-$direction as shown in
Fig.~\ref{fig:1}(a), then the transmitted field reads
  \begin{subequations}\label{eq:5}
  \begin{equation}\label{eq:5.1}
  \eo(\rr)=\eik+\sum_{m} T_{m} e^{i {\bf k}^{+}_{m} \cdot \rr}\,,\ \ \ z>h+R\,,
  \end{equation}
where  ${T}_{m}$ are the amplitudes of the transmitted modes and the first term describes the incident radiation whose amplitude is set to one. Similarly, in the region to the left of the structure, the electric field is given by
  \begin{equation}\label{eq:5.2}
  \eo(\rr)=\eik+\sum_{m} R_{m} e^{i {\bf k}^{-}_{m} \cdot \rr}\,,\ \  \ z<-h-R\,,
  \end{equation}
  \end{subequations}
where  ${R}_{m}$ are the amplitudes of the reflected modes. The choice between $\kk^{+}_{m}$ and $\kk^{-}_{m}$ is dictated by the outgoing wave boundary condition at the spatial infinity $|z|\rightarrow\infty$. In particular, the field corresponding to closed channels in Eqs.(\ref{eq:5}) decays exponentially in the asymptotic region $|z|\rightarrow \infty$. So the closed channels do not contribute to the energy flux carried by the scattered wave. The scattered flux is carried only by the radiation in open diffraction channels. The condition $k^{2}_{m}= (\kx+2\pi m)^2$ defines the {\it threshold} for opening the $m^{th}$ diffraction channel. The coefficients ${R}_{m}$ and ${T}_{m}$ for open channels are the reflection and transmission coefficients, respectively.\\
\indent The transmission and reflection amplitudes may  be inferred from the solution $\eo$ of the Lippmann-Schwinger integral equation \cite{b2}
  \begin{equation}
  \eo(\rr)=\eik +\frac{k^2}{4 \pi}\int \left(\ee(\ro)-1\right)\eo(\ro)G(\rr|\ro)d\ro\,,
  \label{eq:6}
  \end{equation}
in which $\displaystyle{G(\rr|\ro)=i \pi \hz (k |\rr-\ro|)}$ is the 2-dimensional free-space Green's function for the Helmholtz operator $\triangle + k^2$ with outgoing boundary conditions, and $H_{\sz}$ is the Hankel function of the first kind of order $0$. It should be stressed that this equation is to be understood in the distributional sense. This is because the potential under consideration is neither compactly supported nor does it vanish at infinity so that the usual methods that establish this equation cannot be used \cite{b2}. See Appendix~\ref{asec:1} for details.\\
\indent On a different note, observe that, unless there is no incident radiation (the term $\eik$ is omitted), the solutions to Eq.(\ref{eq:2}) cannot be square integrable along the z-axis. By the analogy with quantum scattering theory for radially symmetric potentials, such solutions are in the radiation continuum of the energy spectrum. In quantum theory, solutions to the Schr\"{o}dinger equation are eigenvectors of the energy operator whose spectrum $\eg$ form a subset in the real line. The upper part (positive) of the spectrum  is continuous and corresponds to scattered states that can carry the probability flux to the spatial infinity. Below the continuous part of the spectrum, i.e., $\eg<0$,
the energy spectrum is discrete (see Fig.~\ref{fig:1}(b)). It corresponds to bound states. Bound states have a finite $L_2$ norm (they decay fast enough at the spatial infinity). It was first proved by Von Neumann and Wigner in 1929  that under special circumstances, there might exist bound states in the radiation continuum. A counterintuitive physical peculiarity of such states is that a bound state is a standing wave in a potential well (an attractive potential), while the conventional quantum mechanical wisdom would suggest that for a potential bounded above a standing wave with the energy in the radiation continuum should tunnel through the potential barrier to the spatial infinity and, hence, cannot be stable. Nevertheless, such states do exist and the theory of their formation is now well developed \cite{b4,b5,b7}. The goal here is to establish a similar picture for electromagnetic excitations in the periodic double array of dielectric cylinders, and, specifically, to find the conditions on the physical parameters of this system under which  the bound states in the radiation continuum exist, their eigen-frequencies, and the analytic form of the corresponding electromagnetic fields. The very existence of bound states for this system was first demonstrated by numerical simulations \cite{b12}. Here a complete analytic study of the system is given.

Any bound state is a solution of the generalized eigenvalue problem (\ref{eq:2}) (no incident radiation term) and, hence, is fully characterized by the spectral parameter ${\cal E}=k^2> 0$ and the Bloch phase factor $e^{ik_x}$ because of the boundary condition (\ref{eq:3}). The pair $({\cal E}, e^{ik_x})$ is viewed as a point on the (half) cylinder $\mathbb{R}_{+}\times { S}^1$ where ${\cal E}\in \mathbb{R}_+$ and $e^{ik_x}\in { S}^1$. It will be called a {\it spectral cylinder} (or spectral space) of a periodic grating. In contrast to quantum mechanical bound states in spherically  symmetric systems, the Bloch boundary condition requires a more adequate classification of bound states here. In order to identify bound states in the radiation continuum, one has first to determine the region of the spectral cylinder occupied by the radiation states. In the asymptotic region $|z|\rightarrow\infty$, a harmonic time dependent solution $E=\eo e^{-i \omega t}$ to Eq.(\ref{eq:1}) is characterized by the pair $(\eg,\kx)$ where $\eg=k^2 = \omega^2/c^2$. Given  $\eg$ and $\kx$, the field outside the scattering region is completely determined by its behavior in the diffraction channels as specified in Eqs.(\ref{eq:5}). Consequently, if two radiation modes have the same ${\cal E}$, while their parameters $k_x$ differ by a $2\pi$-multiple, say, by $2\pi m_0$, then they have exactly the same open diffraction channels because in the classification introduced in Eq.(\ref{eq:4}) the difference of channels would merely mean the relabeling $m\rightarrow m-m_0$ (or the same change of the summation index in Eq. (\ref{eq:5})). Therefore the radiation modes correspond to points $({\cal E}, e^{ik_x})$ on the spectral cylinder for which one or more diffraction channels are open.\\
\indent The spectral cylinder can be partitioned into sections associated with a fixed number of open diffraction channels. The diffraction thresholds appear as curves separating these portions of the cylinder. Indeed, let $[k_{x}]$ designate the argument of $e^{i k_{x}}$ in $(-\pi,\pi]$, i.e., $[k_{x}] = k_x \ \mod{2\pi}$. Then, up to the aforementioned reordering,  the diffraction thresholds on the spectral cylinder  are exactly
 \begin{subequations}\label{eq:7}
  \begin{equation}\label{eq:7.1}
  \eg_{\pm n}(k_{x})=(2\pi n \pm |[k_{x}]|)^2
\,, \qquad n=0,1,2,3...\,,
  \end{equation}
and they appear in the order,
  \begin{equation}\label{eq:7.2}
  \eg_{0} \leq \eg_{-1} \leq \eg_{1} \leq \eg_{-2} \leq \eg_{2} \leq \eg_{-3} \leq \eg_{3} \dots
  \end{equation}
  \end{subequations}
\indent
If $k_x$ is identified with the angular variable spanning the
compactified direction of the spectral cylinder, then
the diffraction thresholds are curves in an ever rising order on the spectral cylinder with nodes on the axes $[k_{x}]=0$ and $[k_{x}]=\pi$. The curve $\eg=\eg_{0}(\kx)$ is the threshold below which no radiation modes exist, and therefore, the radiation continuum lies immediately above this curve. This continuum is split into distinct regions corresponding to a fixed number of open channels by consecutive thresholds as indicated by Eq.(\ref{eq:7.2}). These regions will be labeled as radiation continuum I, radiation continuum II, radiation continuum III, etc., where the Roman numeral indicates the number of open channels in each region. See Fig.~\ref{fig:1}(b), (c), and (d).\\
\indent All the solutions to Eq.(\ref{eq:2}) below the threshold $\eg_{0}$, if any, must be bound states, i.e., they decay exponentially in all diffraction channels and, hence, have a finite $L_2$ norm in the space ${ S}^1\times \mathbb{R}$ spanned by $(x,z)$ ($x$ is compactified into a circle ${ S}^1$ because of the boundary condition (\ref{eq:3})). In contrast, radiation modes behave as harmonic functions in the asymptotic region $|z|\rightarrow \infty$ and, hence, do not have a finite $L_2$ norm. So the problem is to find, if any, square integrable solutions on ${ S}^1\times \mathbb{R}$ above the curve ${\cal E}={\cal E}_0$ which are the sought-for bound states in the radiation continuum.

\section{Bound states}
\label{sec:2}
\begin{figure}[h t]
\centering
\includegraphics{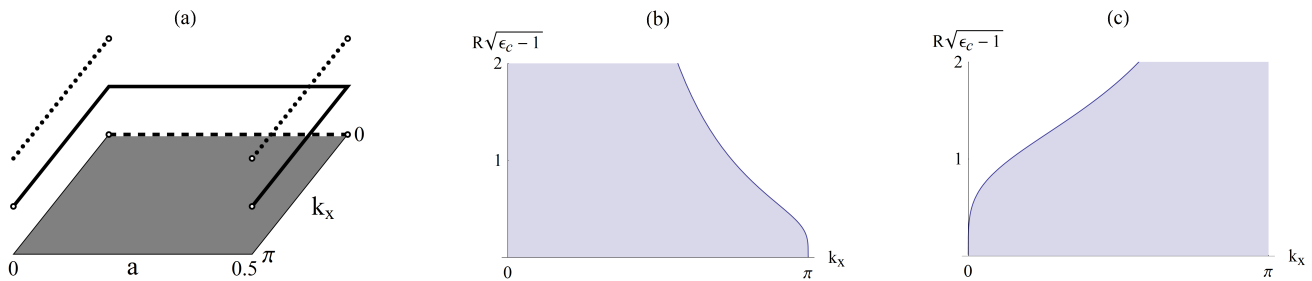}
\caption{Panel (a): Values of the parameters $a$ and $\kx$ for which bound states exist. The bottom level represents the modes below the radiation continuum, the second level represents the modes in the radiation continuum I, and the third level represents the modes in the radiation continuum II.\newline Panels (b) and (c): The shaded areas represents the region of validity of inequalities (\ref{eq:34}) and (\ref{eq:51}), respectively.}
\label{fig:2}
\end{figure}

As defined above, bound states are square integrable solutions of the homogeneous Lippmann-Schwinger integral equation
  \begin{equation}
  \eo(\rr)=\frac{k^2}{4 \pi}\int \left(\ee(\ro)-1\right)\eo(\ro)G(\rr|\ro)d\ro\,, \quad k^2>0\,,
  \label{eq:8}
  \end{equation}
which satisfy Bloch's boundary condition (\ref{eq:3}). The square integrability here means a finite $L_{2}$ norm in ${ S}^{1}\times \mathbb{R}$ (with the x-direction compactified into a circle). As also noted above, Eq.(\ref{eq:7}) is understood in the distributional sense (see Appendix~\ref{asec:1} for details). This is a generalized eigenvalue problem because the Green's function $G(\rr|\ro)$ also depends on the spectral parameter $k^2$. In this problem, the Bloch parameter $\kx$ may be restricted to the interval $[-\pi,\pi]$. It must be stressed that this restriction is only permitted by the absence of an incident wave which otherwise determines the phase factor in (\ref{eq:3}).\\
\indent The task is to determine the values of $a$, $\kx$, $h$ and $k$ that allow for the existence of nontrivial solutions $\eo$ to Eq.(\ref{eq:8}) for fixed radius $R$ and fixed dielectric constant $\eec$ in the limit of
thin cylinders, i.e., $k R\ll 1$. In what follows,
the existence of a bound state always means
the existence of a wavenumber $k$ at which the bound state $\eo$ occurs. The main results established in the present
study may be split into the following cases which are also summarized in Fig.~\ref{fig:2}(a):
\begin{itemize}
\item Below the radiation continuum: Bound states exist for all $\kx$, $a$ and for all distances between the arrays.
\item Continuum I (one open diffraction channel):
Bound states only exist if \begin{itemize}
                                \item either $\kx=0$ and $a\in[0,\sh]$ is arbitrary
                                \item or $a\in \{0,\frac{1}{2}\}$ and $\kx\in (-\pi,\pi)$ is arbitrary
                                \end{itemize}
Under these conditions,
for each pair $(a,\kx)$ there is a discrete set of distances between the arrays at which bound states exist.
\item Continuum II (two open diffraction channels):
Bound states only exist for \\ $a=0$ or $a=\frac{1}{2}$ and for a certain dense set of values of $\kx$.\\ For each allowable pair $(a,\kx)$, there is exactly one or two distances between the arrays at which the bound states exist.
\item Continuum N, N$\geq 3$
(three or more open diffraction channels):
Bound states exist only for specific values of the radius $R$ and the dielectric constant $\eec$.
\end{itemize}

\indent In all the above cases, bound states occur in two types, symmetric and skew-symmetric relative
to the transformation $z\rightarrow -z$ and $x\rightarrow a+x$,
when
the shift parameter takes the boundary values,
$a=0$ or $a=\sh$. However,
this classification cannot be established for double arrays
with intermediate values of the shift parameters,
$a\in (0,\sh)$.\\
\indent The above classification of the bound states in the cases $a=0$ and $a=\sh$ is relevant when analyzing the
(discrete) values
of
the distance $2h$  between the two arrays
at which the bound states occur.
When the double array is symmetric ($a=0$),
then these
values always exist in the whole range of $h\in (R,\infty)$.
On the contrary, skew-symmetric bound states will only occur for those values of $h$ that exceeds a certain minimal threshold, and this
threshold increases as
$R^2(\eec-1)\rightarrow 0$.
In other words, when
the scattering structure becomes more transparent, the two arrays have to be taken further apart
in order for skew-symmetric bound states to form. A similar phenomenon is observed for the skew-symmetric
array ($a=\frac 12$). In this case, however,
the
skew-symmetric modes behave as the symmetric ones
in the previous case and vice versa, i.e., the
minimal threshold distance exists for the symmetric
modes and increases as the system becomes more
transparent ($R^2(\eec-1)\rightarrow 0$), while
the skew-symmetric modes occur at a discrete set
of values of $h$ in the whole range $h\in (R,\infty)$.\\
\indent The existence of the above classification
can be established by studying the symmetry
of the function $\varepsilon(x,z)$. Consider
the operator $P_{a}$ defined by $P_{a}\eo(x,z)=e^{-i a \kx}\eo(x+a,-z)$. For $a=0$ or $a=\sh$, the operator $P_{a}$ commutes with the operator $H$ of Eq.(\ref{eq:2}) and has eigenvalues $\pm 1$ since it is a projection, i.e., $P^{2}_{a}=1$. By virtue of the commutativity,
each bound state $\eo$ is an eigenfunction of $P_{a}$ and, hence,
\begin{equation*}
\eo(x+a,-z)=\pm e^{i a \kx} \eo(x,z)\,.
\end{equation*}
Symmetric states may then be defined as those for which $\eo(x+a,-z)=e^{i a \kx}\eo(x,z)$ while the skew-symmetric states satisfy $\eo(x+a,-z)=-e^{i a\kx}\eo(x,z)$. For $a=0$ this simply means that the first states are even in $z$ while the second are odd.\\
\indent This description cannot be generalized to an arbitrary shift $a$ since $P_{a}$ no longer commutes
with $H$ if $a\in (0,\sh)$. Instead, it follows
from the symmetry of $\varepsilon(x,z)$ that the operator
$Q_{a}\eo(x,z)=\eo(a-x,-z)$ commutes with $H$
for all $a$ and also $Q^{2}_{a}=1$.
However, the operator $Q_{a}$ is not symmetric in the subspace of functions in $L_{2}({S}^{1}\times \mathbb{R})$ satisfying condition (\ref{eq:3}) unless $\kx=0$ or $\kx=\pm\pi$. Indeed if it were, then any bound state $\eo$ would have been
an eigenfunction of $Q_{a}$, and, therefore, $\eo(a-x,-z)=\pm\eo(x,z)$. By replacing $x$ by $x+1$ and applying condition (\ref{eq:3}) one gets $e^{i\kx}\eo(x,z)=e^{-i\kx}\eo(x,z)$ so that $\kx=0$ or $\kx=\pm\pi$.\\
\indent As noted above, the bound states are found by perturbation theory in the limit of thin cylinders. The technical details are given in Appendix~\ref{asec:2} where it is shown that non-trivial solutions to Eq.(\ref{eq:7}) can only exist if the fields $\cph$ and $\cpa$ on the two cylinders positioned at $(0,-h)$ and $(a,h)$ respectively do not simultaneously vanish and satisfy the homogeneous system of equations
\begin{equation}
\begin{cases}
\pst\cpa+\pap\cph=0\\
\pam\cpa+\pst\cph=0
\end{cases}
\label{eq:9}
\end{equation}
in which the coefficients $\pst$ and $\pamp$ are given by the following expressions
\begin{align}
\pst(k,\kx)=&\sum_{m=-\infty}^{\infty}\left(\frac{1}{\kzm}-\frac{1}{2\pi i (|m|+1)}\right)+\frac{i}{2\pi}\left(\frac{1}{\dlo(k)}+2\ln(2\pi R)\right)\nonumber\\
\pamp(a,h,k,\kx)=&\sum_{m=-\infty}^{\infty} \displaystyle{\frac{e^{i\left(\pm a (\kx+2\pi m)+2h\kzm\right)}}{\kzm}}
\label{eq:10}
\end{align}
with $\kzm=\sqrt{k^2-(\kx+2\pi m)^2}$ being the z-component of the wave vector in the $m^{\tx{th}}$ diffraction channel and
\begin{equation}
\dlo(k)= \left(\frac{k R}{2}\right)^2 (\eec-1)
\label{eq:11}
\end{equation}
In particular $\dlo(k)>0$ as $\eec>1$. It should also be kept in mind throughout the work that $\dlo(k)\ll 1$ in the limit
considered $kR\ll 1$. In terms of the fields on the cylinders in Eqs.(\ref{eq:9}), the electric field strength everywhere off the scatterers is then given by,
\begin{equation}
\eo(\rr)=2\pi i \dlo(k)\left(\cpa\sum_{m} \frac{e^{i\left((x-a)(\kx+2\pi m)+|z-h|\kzm\right)}}{\kzm}+\cph \sum_{m} \frac{e^{i\left(x(\kx+2\pi m)+|z+h|\kzm\right)}}{\kzm}\right)
\label{eq:12}
\end{equation}
\indent The system of Eqs.(\ref{eq:9}) admits non-trivial solutions if and only if its determinant
\begin{equation}
\Delta(a,h,k,\kx)=\Phi^{2}_{\sz}-\pap\pam
\label{eq:13}
\end{equation}
vanishes at some point $(a,h,k,\kx)$ in the space of system parameters. This is the condition
for bound states to exist, no matter if they are
below or in the radiation continuum.

In the following two subsections, roots of $\Delta(a,h,k,\kx)$ in $k$ for fixed $a$,$h$, and $\kx$ are analyzed to find bound states below the radiation continuum as well as bound states in the continuum I. The analysis of the higher continua in the spectrum, while being similar to the case of the continuum I, is technically more involved. To avoid excessive technicalities before the discussion of applications of resonances with the vanishing width to a near field amplification, the analysis of higher continua is postponed to Section~\ref{sec:5}. In each study, the spectral parameter $k$ ranges
over an open interval in which
 the functions $\pst$ and $\pamp$ are analytic in $k$ and diverge at the endpoints. Also, note that, since $\Delta(a,h,k,\kx)$ is even in $\kx$, the range of this parameter
can be reduced from $[-\pi,\pi]$ to $[0,\pi]$.
Throughout the rest of this section as well as in Section~\ref{sec:5}, the symbols
 $\php$, $\phc$,  and $\phs$
denote, respectively, the following functions:
\begin{equation}
\php(k,\kx)=\tx{Im}\left(\pst(k,\kx)\right)\,,
\quad \phc(a,h,k,\kx)=\sum_{m^{cl}}
\frac{e^{-2h\qzm}}{\qzm}\cos(2\pi a m)\,, \quad \phs(a,h,k,\kx)=\sum_{m^{cl}}  \frac{e^{-2h\qzm}}{\qzm}\sin(2\pi a m)
\label{eq:14}
\end{equation}
where the superscript "$cl$" in $m^{cl}$ means that the sums are taken over all $m$'s that correspond to closed diffraction channels, and  $\qzm$ is the imaginary part of $\kzm$ when the $m^{\tx{th}}$ channel is closed, i.e., $\qzm=\sqrt{(\kx+2\pi m)^2-k^2}$ if $k^2<(\kx+2\pi m)^2$. Recall that $\kzm=\sqrt{k^2-(\kx+2\pi m)^2}$. By construction,
the functions (\ref{eq:14}) are always real-valued.

\subsection{Bound states below the radiation continuum}
\label{sec:3}

 Bound states below the radiation continuum are nontrivial solutions to the homogeneous Lippmann-Schwinger integral equation when all diffraction channels are closed,
$\kx\in (0,\pi]$ and  $0<k<\kx$. In this case the determinant of Eq.(\ref{eq:13}) factorizes as
\begin{equation}
\Delta(a,h,k,\kx)=-\Psi^{+}\Psi^{-}, \quad \Psi^{\pm}=\php\mp\sqrt{\Phi^{2}_{c}+\Phi^{2}_{s}}
\label{eq:15}
\end{equation}
hence bound states will exist for wavenumbers $k$ at which the functions $\Psi^{+}$ and $\Psi^{-}$ vanish. That such wavenumbers exist for $\Psi^{+}$ follows from the limits
\begin{equation*}
\lim_{k\rightarrow 0^{+}} \Psi^{+}(k)=+\infty\,, \quad \lim_{k\rightarrow \kx^{-}} \Psi^{+}(k)=-\infty
\label{eq:16}
\end{equation*}
and the Intermediate Value Theorem. In particular, for each triplet $(a,h,\kx)$, there exists a wavenumber $k^{+}(a,h,\kx)$ at which $\Psi^{+}(k^{+})=0$. It is shown in Appendix~\ref{asec:4} that in the leading order of $\dlo(\kx)$,
\begin{equation}
k^{+}\approx \kx-\frac{8\pi^2\dlot(\kx)}{\kx}\,.
\label{eq:16.1}
\end{equation}
\indent The function $\Psi^{-}$ can also be shown to have roots when the distance $2h$ is sufficiently large or when $\kx$ is close to $\pi$, this gives a second family of wavenumbers $k^{-}(a,h,\kx)$ at which bound states occur. These assertions result from the limits at $0$ and at $\kx$ of $\Psi^{-}$. These limits are,
\begin{align*}
&\lim_{k\rightarrow 0^{+}} \Psi^{-}(k)=\infty\,, \nonumber\\
&\lim_{k\rightarrow \kx^{-}}\Psi^{-}(k)=-4h \cos^{2}(\pi a)-\frac{\sin^{2}(\pi a)}{\sqrt{\pi(\pi-\kx)}}+\frac{1}{2\pi \dlo(\kx)}+\bigo(1)\,.
\label{eq:17}
\end{align*}
\indent In particular for fixed $\kx\in (0,\pi]$ and $a\neq \sh$, $h$ can be chosen sufficiently large for the last limit to be negative and therefore by the Intermediate Value Theorem a wavenumber $k^{-}$ such that $\Psi^{-}(k^{-})=0$ exists on the interval $(0,\kx)$. Conversely, if $h$ is fixed, then
\begin{equation}
\lim_{\kx\rightarrow \pi^{-}} \lim_{k\rightarrow \kx^{-}} \Psi^{-}=
                            \begin{cases}
                            -\infty &\text{if $a\neq 0$}\\
                            \displaystyle{-4h+\frac{1}{2\pi\dlo(\pi)}}+\bigo(1) &\text{if $a=0$}
                            \end{cases}
                            \label{eq:18}
                            \end{equation}
Thus if $0<a\leq \sh$ and $\kx$ is sufficiently close to $\pi$ then the equation $\Psi^{-}(k)=0$ has a root $k^{-}(a,h,\kx)$. When $a=0$, the sign of the limit in Eq.(\ref{eq:18}) depends on the radius of the scattering cylinders, the dielectric constant $\eec$ and the distance between the two arrays. If $h$ is large enough or $R^2(\eec-1)$ is not too small, then the limit is negative and the existence of the wavenumber $k^{-}$ follows. Because of a complicated dependence of
the existence condition
for the second family of wavenumbers on the physical parameters of the system, such a simple analytic approximation as for $k^{+}$ above is not possible for $k^{-}$.

Finally, from Eqs.(\ref{eq:9}) and $(\ref{eq:12})$
the bound state solutions $E^{\pm}$ at
the wavenumbers $k^{\pm}$ are obtained. As any solution to a homogeneous equation, they can only be determined up to a normalization constant which is chosen to be the value of the electric field $E^{\pm}(-h\ez)$ on the cylinder at $(0,-h)$. In terms of this value, the electric field on the cylinder at $(a,h)$ is then
\begin{equation*}
E^{\pm}(a\ex+h\ez)=\pm e^{i(\phi+a \kx)} E^{\pm}(-h\ez), \quad \phi=\arg\left(\sum_{m} \frac{e^{-2h\qzm+2\pi i a m}}{\qzm}\right)\Bigg|_{k=k^{\pm}}
\label{eq:20}
\end{equation*}
and everywhere off the scattering cylinders it is:
\begin{equation}
E^{\pm}(\rr)=2\pi i \dlo(k)E^{\pm}(-h\ez)\sum_{m}\frac{e^{i x(\kx+2\pi m)}}{\kzm}\left(e^{i|z+h|\kzm}\pm e^{i\left(|z-h|\kzm-2\pi a m +\phi\right)}\right)\Bigg|_{k=k^{\pm}}
\label{eq:21}
\end{equation}

\subsection{Bound states in the radiation continuum I: One open diffraction channel}
\label{sec:4}
\begin{figure}[h t]
    \centering
    \includegraphics[scale=0.9]{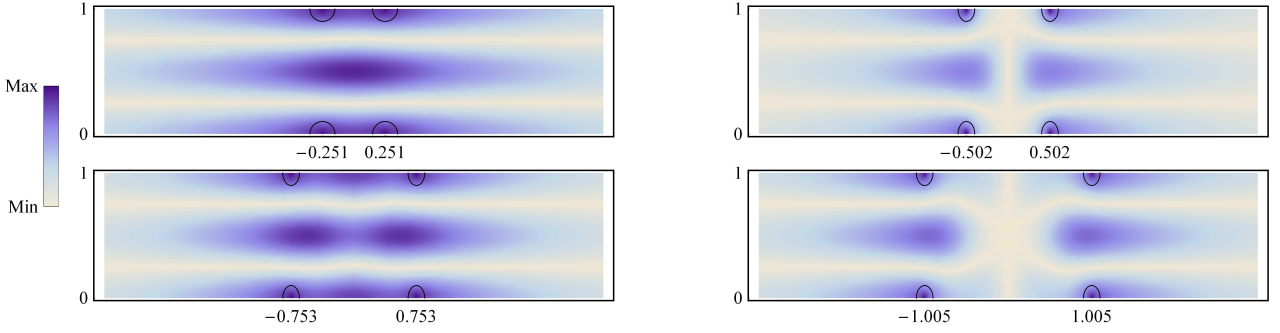}
    \caption{(Color online) The modes $E_{n}$ as defined in Eq.(\ref{eq:39}) for $a=0$, $R=0.1$, $\eec=1.5$, and $\kx=0$. The panels show $E_{n}$ as a function of $x$ (vertical axis) and $z$ (horizontal axis). The color shows the absolute value of $E_{n}$ as indicated on the left inset. The positions of cylinders are indicated by a solid black curve. The values of $h_{n}$ are shown below each panel. Top left: $E_{1}$, the symmetric mode for the smallest $h=h_{1}$. Top right: $E_{2}$, the skew-symmetric mode at $h=h_{2}$. Bottom left: $E_{3}$, the second symmetric mode at $h=h_{3}$. Bottom right: $E_{4}$, the second skew-symmetric mode at $h=h_{4}$.}
    \label{fig:3}
    \end{figure}

 When only the $0^{\text{th}}$-order diffraction channel is open, $\kx\in [0,\pi)$ and $\kx<k<2\pi-\kx$, the determinant of Eq.(\ref{eq:13}) can be rewritten in the
following form convenient for the analysis
\begin{equation}
\Delta(a,h,k,\kx)=\frac{\sin^2(2h\kz)}{k^{2}_{z}}+\Phi^{2}_{s}-\Psi^{+}\Psi^{-}+ \frac{2i}{\kz}\left(\Psi^{+}\sin^2(h\kz)+\Psi^{-}\cos^2(h\kz)\right)
\label{eq:22}
\end{equation}
where $\Psi^{\pm}(a,h,k,\kx)=\php(k,\kx)\pm\left(\frac{\sin(2h\kz)}{\kz}-\phc(a,h,k,\kx)\right)$ and the functions $\php$,$\phc$ and $\phs$ are given in Eqs.(\ref{eq:14}). Thus,
bound state exist if both the real and imaginary parts
of (\ref{eq:22}) vanish:
\begin{equation}
\begin{cases}
\displaystyle{\frac{\sin^2(2h\kz)}{k^{2}_{z}}+\Phi^{2}_{s}}=\Psi^{+}\Psi^{-}\vspace{0.1 cm}\\
\Psi^{+}\sin^2(h\kz)+\Psi^{-}\cos^2(h\kz)=0
\end{cases}
\label{eq:23}
\end{equation}
The first of these equations implies that $\Psi^{+}\Psi^{-}\geq 0$. If this inequality were to be strict, then the second equation would not have held, and, therefore,
$\Psi^{+}\Psi^{-}=0$. Thus,  the first equation implies
that $\phs=0$ and $\sin(2h\kz)=0$. In turn,
the latter equation implies that either $\cos(h\kz)=0$ or $\sin(h\kz)=0$, and, therefore, the system of Eqs.(\ref{eq:23}) splits into two systems, namely,
\begin{equation}
\begin{cases}
\phs(a,h,k,\kx)=0\\
\cos(h\kz)=0\\
\Psi^{+}(a,h,k,\kx)=0
\end{cases}
\qquad
\begin{cases}
\phs(a,h,k,\kx)=0\\
\sin(h\kz)=0\\
\Psi^{-}(a,h,k,\kx)=0
\end{cases}
\label{eq:24}
\end{equation}
To solve the first equation of each system,
the series for $\phs$ is rewritten as,
\begin{equation}
\phs=-\sum_{m=1}^{\infty} c_{m}\sin(2\pi a m), \quad c_{m}=\frac{e^{-2h q_{z,-m}}}{q_{z,-m}}-\frac{e^{-2h q_{z,m}}}{q_{z,m}}
\label{eq:25}
\end{equation}
Recall that $\qzm=\sqrt{(\kx+2\pi m)^2-k^2}$. In particular, if $\kx=0$ then $c_{m}=0,\forall m=1,2,3...$ and the equation $\phs=0$ holds trivially. Similarly, if $a=0$ or $a=\sh$, then the equation holds trivially as $\sin(2\pi a m)=0,\forall m$. It turns out that these are the only possible roots of $\phs$ if $h> \frac{\ln 2}{4\pi}\approx 0.055$. This conclusion stems from the following factorization of $\phs$,
\begin{equation}
\phs(a,h,k,\kx)=-\sin(2\pi a)\sum_{m=1}^{\infty}\left[\left(c_{m}-2c_{m+1}+2\sum_{n=1}^{\infty}\left(c_{m+2n}-c_{m+2n+1}\right)\right)\frac{\sin^2(\pi a m)}{\sin^2(\pi a)}\right]
\label{eq:26}
\end{equation}
together with the facts that for $\kx \neq 0$,
\begin{equation}
c_{m}>0 \quad \text{and} \quad \sup_{m=1}^{\infty}\left\{\frac{c_{m+1}}{c_{m}}\right\}=e^{-4\pi h}
\label{eq:27}
\end{equation}
These statements are established in Appendix~\ref{asec:3}. When $h>\frac{\ln 2}{4\pi}$, then $c_{m+1}<\sh c_{m},\forall m$ and therefore each of the summands in the series of Eq.(\ref{eq:26}) is nonnegative. Since the first term of the said series does not vanish, it follows that the series does not vanish. Consequently, if $\kx\neq 0$ then $\phs=0$ if and only if $\sin(2\pi a)=0$ ,i.e., $a=0$ or $a=\sh$.\\
\indent It is remarkable that under
the restriction $h>\frac{\ln 2}{4\pi}$ no solution
to Eqs.(\ref{eq:24}) is lost. Indeed, since $\sin(2h\kz)=0$ at a bound state, it follows that
\begin{equation*}
h=\frac{n\pi}{2\kz}, \quad \kz=\sqrt{k^2-k^{2}_{x}}
\label{eq:28}
\end{equation*}
for some positive integer $n$. In the ranges considered, $\kz<2\pi$ and therefore $h>\frac{n}{4}\geq  \frac{1}{4}>\frac{\ln 2}{4\pi}$. Thus a necessary condition for the existence of bound states is that either $\kx=0$ while $a\in [0,\sh]$ is arbitrary or $a$ is either $0$ or $\sh$ while $\kx\in [0,\pi)$ is arbitrary. In set notation,
\begin{equation}
(a,\kx)\in L=\Big([0,\sh]\times \{0\}\Big) \cup \Big(\{0,\sh\}\times [0,\pi)\Big)
\label{eq:29}
\end{equation}
The set $L$ is represented by the second level of Fig.\ref{fig:2}(a).\\
\indent Let us turn to solving the last two equations in each of the systems in Eqs.(\ref{eq:24}). For this purpose, the function $\Psi_{n}$ is defined for each positive integer
$n$ by,
\begin{equation}
\begin{split}
\Psi_{n}(k,\kx,a)&=\begin{cases}
                    \displaystyle{\Psi^{+}(a,\frac{n\pi}{2\kz},k,\kx)} \quad \text{if $n$ is odd}\vspace{2 mm}\\
                    \displaystyle{\Psi^{-}(a,\frac{n\pi}{2\kz},k,\kx)} \quad \text{if $n$ is even}
                    \end{cases}\\
                    &=\frac{1}{2\pi\dlo(k)}+\sum_{m\neq 0}\left(\frac{1}{2\pi(|m|+1)}-\frac{1-(-1)^{n}\cos(2\pi a m)e^{-n\pi \qzm k^{-1}_{z} }}{\qzm}\right)+\frac{1}{\pi}\left(\frac{1}{2}+\ln(2\pi R)\right)
                    \end{split}
\label{eq:30}
\end{equation}
where $k\in (\kx,2\pi-\kx)$. Then the systems of Eqs.(\ref{eq:24}) split into the countable set of systems
\begin{equation}
\begin{cases}
h=\displaystyle{\frac{n\pi}{2\sqrt{k^2-k^{2}_{x}}}}\\
\Psi_{n}(k,\kx,a)=0
\end{cases}\qquad n=1,2,3...
\label{eq:31}
\end{equation}
where the systems corresponding to odd $n$ arise from the first of systems (\ref{eq:24}) and those corresponding to even $n$ result from the second system. In each of the systems (\ref{eq:31}), the second equation determines the wavenumbers at which bound states occur. In turn, by the first equation,
these wavenumbers determine the distances $h$ that allow for the bound states to exist. In the next paragraph it is shown that, for fixed $(a,\kx)\in L$, each system can admit at most one solution. Hence the set of distances $h$ allowing the existence of bound states is discrete.\\
\indent Let $(a,\kx)\in L$ be fixed. It is shown in Appendix~\ref{asec:3} that the function $k\mapsto\Psi_{n}$ is monotone decreasing on its domain $(\kx,2\pi-\kx)$ and therefore admits atmost one root in the said domain. Moreover, the root only exists if the limits of $k\mapsto\Psi_{n}$ at $\kx$ and $2\pi-\kx$ are of opposite sign. Specifically, the limit at $\kx$ must be positive while the limit at $2\pi-\kx$ must be negative. The first limit is,
\begin{equation*}
\lim_{k\rightarrow k^{+}_{x}} \Psi_{n}=\begin{cases}
                                     \infty \quad \text{if $\kx=0$}\\
                                     \displaystyle{\frac{1}{2\pi\dlo(\kx)}+\sum_{m\neq 0}\left(\frac{1}{2\pi(|m|+1)}-\frac{1}{\sqrt{4\pi^2m^2+4\pi m\kx}}\right)+\frac{1}{\pi}\left(\frac{1}{2}+\ln(2\pi R)\right)} \quad \text{if $\kx\neq 0$}
                                     \end{cases}
                                     \label{eq:32}
                                     \end{equation*}
 The requirement that this limit be positive when $\kx\neq 0$ puts a restriction on the values of $R$ and $\eec$ that allow for the existence of bound states. However, this is a too complicated condition to analyze. A weaker, but easier to analyze, condition is obtained by first rewriting the positivity condition as,
 \begin{equation}
 \frac{2}{\pi R^2(\eec-1)k^{2}_{x}}>\sum_{m=1}^{\infty}\left(\frac{1}{\sqrt{4\pi^2m^2-4\pi m\kx}}+\frac{1}{\sqrt{4\pi^2m^2+4\pi m\kx}}-\frac{1}{\pi m}\right)+\frac{1}{\pi}\left(\frac{1}{2}-\ln(2\pi R)\right)
 \label{eq:33}
 \end{equation}
 The rearrangement of the series is made to ensure that all the summands in the series are nonnegative; they vanish at $\kx=0$. Also, since the cylinders are thin, it may be assumed that $R<\frac{\sqrt{e}}{2\pi}\approx 0.262$ so that $\frac{1}{2}-\ln(2\pi R)>0$. Therefore all the summands in Eq.(\ref{eq:33}) are nonnegative and hence the left hand side must be larger than each of the summands on the right individually. To estimate the threshold value of $R\sqrt{\eec-1}$ below which bound states may exist, the first term ($m=1$) is retained in the sum (\ref{eq:33}). This term is then written in the form $k_{x}^{2}(\pi-\kx)^{-1/2}g(\kx)$ to isolate its branch point at $\kx=\pi$ and its root at $\kx=0$ which is of multiplicity 2. The minimization of $g(\kx)$ on $[0,\pi]$ produces an estimate:
 \begin{equation}
 R\sqrt{\eec-1}<\frac{C\sqrt[4]{\pi-\kx}}{k^{2}_{x}}
 \label{eq:34}
 \end{equation}
 where
 \begin{equation*}
 C=\pi^{\frac{3}{4}}\sqrt{2}\left(\min_{0\leq t\leq 1}\frac{1+\sqrt{1+t}+\sqrt{1-t}}{\sqrt{1+t}\left(1+\sqrt{1-t^2}\right)\left(2+\sqrt{1+t}+\sqrt{1-t}\right)}\right)^{-\frac{1}{2}}\approx 5.846
 \label{eq:35}
 \end{equation*}to observe
 In particular when $\kx$ is close to $\pi$, the quantity $R\sqrt{\eec-1}$ must be small enough in order for bound states to exist at all. Figure~\ref{fig:2}(b) shows the regions in which Eq.(\ref{eq:34}) is valid.\\
 \indent As specified above, for bound states to exist
it is required that the limit of $\Psi_{n}$ at $2\pi-\kx$ be negative. This limit is,
 \begin{equation*}
\lim_{k\rightarrow (2\pi-\kx)^{-}} \Psi_{n}=\begin{cases}
                                    -\infty \quad \text{if $n$ is odd and $a\neq \frac{1}{2}$ or $n$ is even and $a\neq 0$}\vspace{2 mm}\\
                                    \displaystyle{-\frac{n\pi}{\sqrt{\pi(\pi-\kx)}}+\frac{1}{2\pi\dlo(2\pi-\kx)}+\bigo(1)}\quad\text{otherwise}
                                    \end{cases}
                                    \label{eq:36}
                                    \end{equation*}
 In particular, the limit is negative except possibly when $n$ is odd and $a=\sh$ or $n$ is even and $a=0$. Even in the latter cases however, the negativity condition may be ensured by taking $\kx$ sufficiently close to $\pi$ so that $k^{-1}_{z}$ is large or by choosing $n$ sufficiently large. Thus, if the parameters $R$ and $\eec$ of the scattering cylinders verify condition (\ref{eq:33}), bound states in the continuum I do exist. To be precise, given a positive integer $n$; then
 \begin{itemize}
 \item $\forall (a,\kx)\in \Big((0,\frac{1}{2})\times \{0\}\Big)\cup \Big(\{0\}\times [0,\pi)\Big)$ there exists a bound state at the wavenumber $k_{2n-1}(a,\kx)\in (\kx,2\pi-\kx)$ and at the distance $2h_{2n-1}(a,\kx)=\frac{(2n-1)\pi}{\sqrt{k^{2}_{2n-1}-k^{2}_{x}}}$ between the two arrays of cylinders. When $a=\sh$, then the wavenumber $k_{2n-1}$ exists for sufficiently large $n$ or for $\kx$ sufficiently close to $\pi$.
 \item $\forall (a,\kx)\in \Big((0,\frac{1}{2})\times \{0\}\Big)\cup \Big(\{\frac{1}{2}\}\times (0,\pi)\Big)$ there exists a bound state at the wavenumber $k_{2n}(a,\kx)\in (\kx,2\pi-\kx)$ and at the distance $2h_{2n}(a,\kx)=\frac{2n\pi}{\sqrt{k^{2}_{2n}-k^{2}_{x}}}$ between the two arrays of cylinders. When $a=0$, then the wavenumber $k_{2n}$ exists for sufficiently large $n$ or for $\kx$ sufficiently close to $\pi$.
\end{itemize}
\indent In the limit of the thin cylinders considered, approximate values can be inferred for the wavenumbers $k_{n}$, $n=1,2,3..$, by only keeping the leading terms in the equations $\Psi_{n}(k)=0$. This is detailed in Appendix~\ref{asec:4}. For instance if $n$ is odd and $a=0$ or $n$ is even and $a= \sh$, then the wavenumber $k_{n}(\kx)$ and the distance $h_{n}(\kx)$ are approximated in the leading order of $\dlo$ by,
\begin{equation}
k_{n}(\kx)\approx 2\pi-\kx-\frac{8\pi^2\dlot(2\pi-\kx)}{2\pi-\kx}\qquad h_{n}(\kx)\approx\frac{n\pi}{4\sqrt{\pi(\pi-\kx)}}\left(1+\frac{2\pi\delta^{2}_{\sz}(2\pi-\kx)}{\pi-\kx}\right)
                    \label{eq:37}
                    \end{equation}
\indent Finally, from Eqs.(\ref{eq:9}) and (\ref{eq:12})
the explicit form the electric field for
the bound states $\{E_{n}\}_{n=1}^{\infty}$
is obtained
at the wavenumbers $\{k_{n}\}_{n=1}^{\infty}$ and the distances $\{2h_{n}\}_{n=1}^{\infty}$ between the arrays of cylinders. The eigenfunction of bound states in the continuum can only be determined up to a multiplicative constant which is chosen to be the value of the electric field $E_{n}(-h_{n}\ez)$ on the cylinder at $(0,-h_{n})$. In terms of this value, the electric field on the cylinder at $(a,h_{n})$ is then,
\begin{equation*}
E_{n}(a\ex+h_{n}\ez)=(-1)^{n+1}e^{i a \kx}E_{n}(-h_{n}\ez)
\label{eq:38}
\end{equation*}
and everywhere off the scattering cylinders it is:
\begin{equation}
E_{n}(\rr)=2\pi i \dlo(k_{n})E_{n}(-h_{n}\ez)\sum_{m}\frac{e^{i x(\kx+2\pi m)}}{k^{n}_{z,m}}\left(e^{i|z+h_{n}|k^{n}_{z,m}}+(-1)^{n+1}e^{i\left(|z-h_{n}|k^{n}_{z,m}-2\pi a m\right)}\right)
\label{eq:39}
\end{equation}
where $k^{n}_{z,m}=\sqrt{k^{2}_{n}-(\kx+2\pi m)^2}$. It can be verified easily that outside the scattering region, i.e., $|z|>h_{n}$, there is no contribution to the field $E_{n}$ from the $0-$order diffraction channel. With this channel being the only open channel, it follows that $E_{n}$ decays exponentially in the asymptotic region $|z|\rightarrow \infty$ and therefore it is square integrable on ${S}^{1}\times {R}$ as required. Figure~\ref{fig:3} shows examples of plots of the absolute values of the fields $E_{n}$.

\subsection{Application: Zero width resonances and near field amplification}
\label{sec:7}

\indent The study of Section~\ref{sec:2} shows that, given two parallel arrays of subwavelength dielectric cylinders of radius $R$ and dielectric constant $\eec$, there are points $(a,\kx)$ for which a bound state exists if the distance between the arrays attains a specific value $2h_{b}$; this value also determines the wavenumber $k_{b}$ of the bound state. Throughout the following discussion, the pair $(a,\kx)$ is fixed.

Consider the scattering problem for the double
array when $h=h_b$. If the incident wave has the
wavenumber $k_{b}$, then the solution to Eq.(\ref{eq:2}) is not unique as any solution of the homogeneous equation
can always be added. The latter are the bound states.
Of course, this ambiguity is related to the fact that
the incident radiation cannot excite the bound state
(which is a wave-guiding mode propagating along
the array), and, hence, an addition condition must be imposed
that the bound state is not present in the system at the
very beginning if one wishes to have a unique solution.
On the other hand, the presence of a bound state
has no effect whatsoever on the scattering amplitudes
as they are defined in the far field zone ($z\rightarrow
\pm \infty$) to which the bound state gives no contribution
anyway.
 This degeneracy disappears as soon as $(h,k) \neq (h_{b},k_{b})$. In the latter case, the solution $\eo$ to Maxwell's equations is uniquely determined by Eq.(\ref{eq:6}). Such irregularity suggests that, as a function of $h$ and $k$, the field $\eo$ is not analytic in the vicinity of the points $(h_{b},k_{b})$. This is indeed the case. It is shown in what follows shortly that the values of the reflected flux as well as those of the fields inside the cylinders near the points $(h_{b},k_{b})$ depend on the path along which these points are approached in the $h,k-$plane. From a mathematical point view, this lack of analyticity is explained by the presence of simple poles at the wavenumbers $k_{b}$ in the field $\eo$ when considered as a function of $k$ for fixed $h=h_{b}$. The objective here is to exploit the existence of these poles to show that the
evanescent field in the scattering problem is
 amplified as compared to the amplitude of the
incident field when $(h,k)$ is close to $(h_b,k_b)$
in some regions of the array, in particular, on the
cylinders.
The effect can therefore be used to
amplify optical non-linear effects in the structure
in a controllable way.
 This is illustrated with an example
of one open channel for an array without the shift, i.e., $a=0$.

\begin{figure}[h t]
    \centering
    \includegraphics{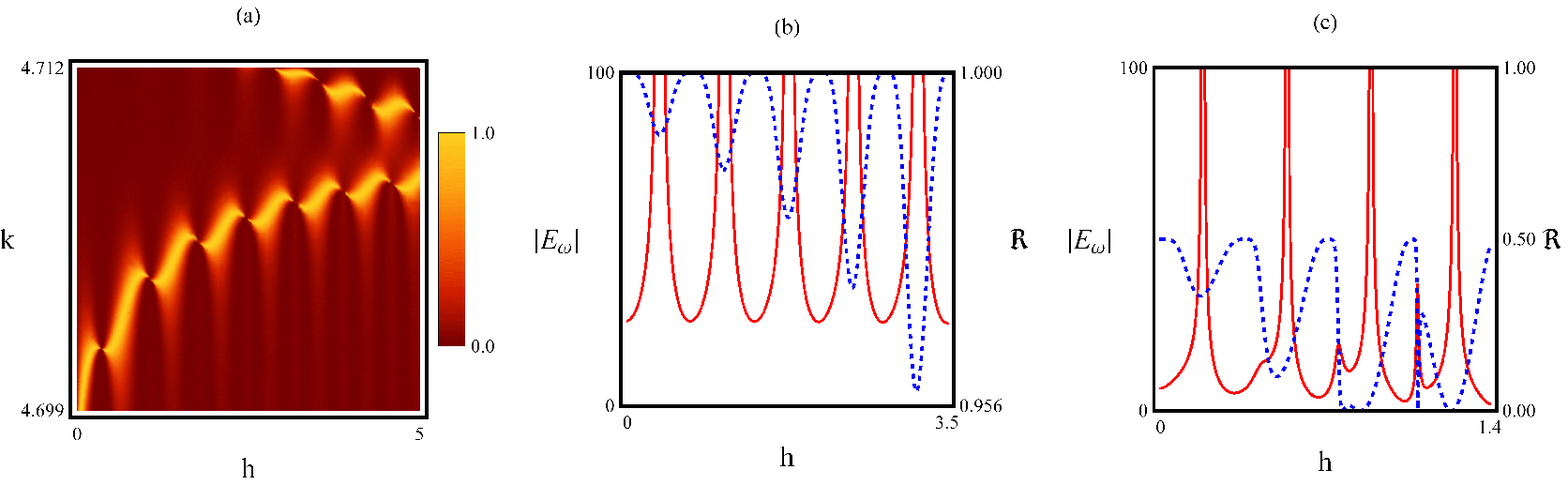}
    \caption{(Color online) The specular coefficient and the electric field on the cylinders near a bound state in the continuum for $a=0$, $R=0.1$, and $\eec=1.5$.\newline Panel (a): Shows the specular reflection coefficient as a function of $h$ and the wavenumber $k$. It is plotted for $\kx=\frac{\pi}{5}$ near the threshold $k_{-1}=\frac{9\pi}{5}$. The specular coefficient is very close to its maximum along the curves $\Psi^{\pm}(h,k)=0$ that determine the resonance positions. The lower (upper) bright region roughly corresponds to the curve $\Psi^{+}(h,k)=0$ ($\Psi^{-}(h,k)=0$). The bound states correspond to the points separating two consecutive bright strips. At the bound states the specular coefficient lacks analyticity and its value depends strongly on how the bound state is approached in the $(h,k)$-plane.\newline Panel (b): The specular coefficient $\rfr(k(h),h)$ (dashed blue curve) and the absolute value of the electric field $\eo(h\ez)(k(h),h)$ (solid red curve) on the cylinder at $(x,z)=(0,h)$ of the double array where $k=k(h)$ is implicitly defined by $\Psi^{+}(h,k)=0$. Along this curve, the electric field diverges near the bound states. \newline Panel (c): Same legend as for Panel (b) in the case of two open channels and $\kx=\pi$. In this case too, the fields on the cylinders diverge in the vicinity of the bound states.}
    \label{fig:4}
    \end{figure}

Suppose that a plane wave of wavenumber $k\in(\kx,2\pi-\kx)$ and unit amplitude impinges the double array. In this case, the {\it specular reflection coefficient} which is the ratio of the reflected flux to the incident flux at the spatial infinity is,
 \begin{equation*}
 \rfr=|R_{\sz}|^2
 \label{eq:77}
 \end{equation*}
where $R_{\sz}$ is the reflection coefficient of the only open diffraction channel, namely, the $0$ order channel as given in Eq.(\ref{eq:5.2}). If $(h,k)$ is not one of the points $(h_{b},k_{b})$, then the reflection coefficient $R_{\sz}$ and the fields inside the cylinders are,
\begin{subequations}\label{eq:78}
 \begin{gather}
 R_{\sz}=-\frac{\cos^2 (h\kz)}{\cos^2 (h\kz)+\frac{1}{2}i\kz \Psi^{+}}+\frac{\sin^2 (h\kz)}{\sin^2 (h\kz)+\frac{1}{2}i\kz \Psi^{-}}\label{eq:78.1}\\
 \cpm=\frac{i\kz}{2\pi\delta_{_{\sz}}(k)}\left(\frac{\cos (h\kz)}{\cos^2 (h\kz)+\frac{1}{2}i\kz \Psi^{+}}\pm i \frac{\sin(h\kz)}{\sin^2(h\kz)+\frac{1}{2}i\kz\Psi^{-}}\right) \label{eq:78.2}
 \end{gather}
 \end{subequations}
 for the functions $\Psi^{\pm}$ of Eq.(\ref{eq:22}) (See Appendix~\ref{asec:2} for details on the derivations of the above expressions). The denominators in both expressions are factors of the determinant $\Delta(0,h,\kx,k)$in Eq.(\ref{eq:22}) (Here $a=0$). In particular, the points $(h_{b},k_{b})$ are roots of the denominators to the specular coefficient and the fields inside the cylinders.\\
 \indent For the illustration purpose,  the behavior of the specular coefficient and the fields inside the cylinders
are studied
as $(h,k)$ approaches a critical point $(h_{b},k_{b})=(h_{2n-1},k_{2n-1})$ for some positive integer $n$ in the $h,k-$plane. As described in Subsection~\ref{sec:4}; at the point $(h_{2n-1},k_{2n-1})$ the following system holds,
 \begin{equation*}
 \begin{cases}
 \cos(h\kz)=0\\
 \Psi^{+}(h,k)=0
 \end{cases}
 \label{eq:79}
 \end{equation*}
 In the rest, the curves $\cos(h\kz)=0$ and $\Psi^{+}(h,k)=0$ will be denoted by $\ccc$ and $\ccp$ respectively and their intersection points, i.e., the points $(h_{2n-1},k_{2n-1})$, will be denoted by $\pn$.\\
 \indent The first observation is that as $(h,k)$ approaches $\pn$, then $\cos(h\kz)\rightarrow 0$ and $\Psi^{+}(h,k)\rightarrow 0$ independently as the curves $\ccc$ and $\ccp$ intersect at a nonzero angle at $\pn$. This may be established through the linearizations of the functions $(h,k)\mapsto \cos(h\kz)$ and $(h,k)\mapsto \kz\Psi^{+}(h,k)$ at $(h_{2n-1},k_{2n-1})$. If $\Delta h=h-h_{2n-1}$ and $\Delta k=k-k_{2n-1}$, then in the vicinity of $\pn$,
 \begin{equation*}
 \begin{split}
 \cos(h\kz)&\approx\xi(\Delta h,\Delta k)=(-1)^n\left(\frac{h_{2n-1} k_{2n-1}}{k_{z,2n-1}}\Delta k+k_{z,2n-1} \Delta h\right)\\
 \frac{1}{2}\kz\Psi^{+}(h,k)&\approx\eta(\Delta h, \Delta k)=\frac{1}{2}k_{z,2n-1}\Big(\partial_{k} \Psi^{+}(h_{2n-1},k_{2n-1})\Delta k+\partial_{h} \Psi^{+}(h_{2n-1},k_{2n-1})\Delta h\Big)
 \end{split}
 \label{eq:80}
 \end{equation*}
 where $k_{z,2n-1}=\sqrt{k^{2}_{2n-1}-k^{2}_{x}}$. The functions $\xi$ and $\eta$ are then linearly independent if,
 \begin{equation*}
 \frac{h_{2n-1} k_{2n-1}}{k_{z,2n-1}}\partial_{h} \Psi^{+}(h_{2n-1},k_{2n-1})-k_{z,2n-1}\partial_{k} \Psi^{+}(h_{2n-1},k_{2n-1})\neq 0
 \label{eq:81}
 \end{equation*}
 That this condition indeed holds can be proved
by examining
the function $\Psi_{2n-1}(k)=\Psi^{+}\left(\frac{(2n-1)\pi}{2\kz},k\right)$ introduced in Eq.(\ref{eq:30}). In Appendix~\ref{asec:3} it is shown that $\partial_{k} \Psi_{2n-1}(k)<0$ for all $k\in (\kx,2\pi-\kx)$. Consequently,
 \begin{equation*}
\partial_{k} \Psi_{2n-1} (k_{2n-1})=-\frac{h_{2n-1}k_{2n-1}}{k^{2}_{z,2n-1}} \partial_{h} \Psi^{+}(h_{2n-1},k_{2n-1})+\partial_{k} \Psi^{+}(h_{2n-1},k_{2n-1})<0
 \label{eq:82}
 \end{equation*}
 This establishes the linear independence of $\xi$ and $\eta$. Thus, as $(\Delta h,\Delta k)\rightarrow (0,0)$,
there should be
$\xi\rightarrow 0$ and $\eta\rightarrow 0$ independently. In the vicinity of the critical point $\pn$ the principal parts of $R_{\sz}$ and $\cpm$ are then,
 \begin{subequations}\label{eq:83}
 \begin{equation}\label{eq:83.1}
 R_{\sz}(h,k)\approx\frac{1}{1+i k_{z,2n-1} \Psi^{-}(h_{2n-1},k_{2n-1})}+\frac{\xi^2( \Delta h,\Delta k)}{\xi^2( \Delta h,\Delta k)+i \eta(\Delta h,\Delta k)}
 \end{equation}
 \begin{equation}\label{eq:83.2}
 \cpm\approx \frac{i k_{z,2n-1}}{2\pi \dlo(k_{2n-1})}\left(\pm i\frac{(-1)^{n+1}}{1+i k_{z,2n-1}\Psi^{-}(h_{2n-1},k_{2n-1})}+\frac{\xi( \Delta h,\Delta k)}{\xi^2( \Delta h,\Delta k)+i \eta(\Delta h,\Delta k)}\right)
 \end{equation}
 \end{subequations}
 The first summands in each of these equations are constant and obey the estimate,
 \begin{equation*}
 \frac{1}{1+i k_{z,2n-1} \Psi^{-}(h_{2n-1},k_{2n-1})}\sim \frac{\dlo(k_{2n-1})}{k_{z,2n-1}}
 \label{eq:84}
 \end{equation*}
 The second summands in Eqs.(\ref{eq:83}) account for the lack of analyticity of the specular coefficient $\rfr$ and the fields $\cpm$ in the vicinity of the critical point $\pn$. In particular, since $\eta\equiv 0$ along the tangent line to $\ccp$ at $\pn$, it follows that along this tangent line and hence along the curve $\ccp$,
 \begin{equation*}
 \cpm\approx \frac{(-1)^{n} i}{2\pi\dlo(k^{2}_{2n-1})\left(1-\frac{h_{2n-1}k_{2n-1}}{k^{2}_{z,2n-1}}\frac{\partial_{h}\Psi^{+}}{\partial_{k} \Psi^{-}}\right)\Delta h}, \quad \Delta h=h-h_{2n-1}\rightarrow 0
 \label{eq:85}
 \end{equation*}
 Thus the electric field inside the cylinders diverges at the points $\pn$. \\
 \indent For the specular coefficient, the significance of the points $\pn$ is that they are positions of resonances with the vanishing width, which, in turn, demonstrates that the
bound state in the radiation continuum are interpreted
as resonances with the vanishing width in the formal
scattering theory.
Indeed, if $h\neq h_{2n-1}$ is fixed and $k_{r}(h)$ is a wavenumber such that $\Psi^{+}(h,k_{r}(h))=0$, then $k_{r}$ is a resonant wavenumber for the specular coefficient $\rfr$. For $k$ near $k_{r}$, the Breit-Wigner theory asserts that $R_{\sz}$ will have the form,
 \begin{equation*}
 R_{\sz}\sim \frac{i\Gamma}{k-k_{r}+i \Gamma}
 \label{eq:86}
 \end{equation*}
 where $2\Gamma$ is the resonance width of the Lorentzian profile of $\rfr=|R_{\sz}|^2$. The half-width $\Gamma$ may be found by expanding the function $k\mapsto \Psi^{+}(h,k)$ in a Taylor series at the resonant wavenumber $k_{r}$, it is,
 \begin{equation*}
 \Gamma=-\frac{2\cos^{2}(h \kz)}{\kz \partial_{k} \Psi^{+}}\Bigg|_{k=k_{r}}
 \label{eq:87}
 \end{equation*}
 so that at the points $\pn$ this width vanishes. Figure~\ref{fig:4} shows plots of the specular coefficients and the electric field along the curve $\ccp$.

\section{Bound states in the radiation continuum N, N$\geq 2$}
\label{sec:5}

When more than one diffraction channel are open, the bound states may still be shown to exist. However they become rarer as the number of diffraction channels increases. The physical reason for that is simple. As suggested in the introduction, a bound state is formed due to a destructive interference of the decay radiation of two quasi-stationary electromagnetic modes localized in the vicinity of each array. If more than one decay channel are open for these modes, then the destructive interference must occur in all the decay channels in order for a bound state to form, which puts more restrictions on the system parameters. Indeed, consider, for instance, possible choices of the parameters $a$ and $\kx$ that allow for the existence of bound states. It was shown that bound states do exist below the radiation continuum for all pairs $(a,\kx)$ ,i.e., no restrictions at all. When one diffraction channel is open, then bound states can form when the pairs $(a,\kx)$ lie on the set $L$ defined in Eq.(\ref{eq:29}). When two diffraction channels are open, then the bound states are shown to only occur if the shift $a$ is $0$ or $\sh$ and for a specific dense set of values of $\kx$. As one goes on to higher levels of the spectral cylinder in Fig.\ref{fig:1}(d), the values of $\kx$ at which bound states may exist become more sparse and are determined by solutions of a system of diophantine equations. The conditions under which bound states exist are formulated first for the studied system in the continuum $N\geq 2$.

The above assertions  follow from the observations that, if the diffraction channels $0\,, -1\,, 1\,,...$ are open, then at a bound state, the parameters $\kx\,, k\,, h\,, a\,, R\,,$ and $\eec$ satisfy the relations,
\begin{equation}
\begin{cases}
\Delta(a,h,\kx,k)=0\\
\sin(2h \kz)=0\\
\sin(2h k_{z,-1})=0\\
\sin(2h k_{z,1})=0\\
\qquad \quad\vdots
\end{cases}
\label{eq:69}
\end{equation}
where $\Delta(a,h,\kx,k)$ is the determinant of Eq.(\ref{eq:13}). The additional equations are a result of the square integrability of the bound states on ${S}^{1}\times \mathbb{R}$. Indeed, if a solution $\eo$ to Eq.(\ref{eq:8}) is to be square integrable on ${S}^{1}\times \mathbb{R}$ then, the function
\begin{equation*}
z\mapsto\int_{0}^{1} |\eo(x,z)|^{2} dx
\label{eq:70}
\end{equation*}
is integrable in $z$ over $\mathbb{R}$. Now,
\begin{equation*}
\int_{0}^{1} |\eo(x,z)|^{2} dx \rightarrow4\pi^2\dlot(k)\begin{cases}
\displaystyle{\sum_{m^{op}}\frac{1}{k^{2}_{z,m}}\left|e^{i h\kzm}\cph+e^{-i (a(\kx+2\pi m)+h\kzm)}\cpa\right|^{2}}\,, &z\rightarrow\infty\vspace{2 mm}\\
\displaystyle{\sum_{m^{op}}\frac{1}{k^{2}_{z,m}}\left|e^{-i h\kzm}\cph+e^{-i( a(\kx+2\pi m)-h\kzm)}\cpa\right|^{2}}\,, &z\rightarrow-\infty
\end{cases}
\label{eq:71}
\end{equation*}
where the superscript "$op$" in $m^{op}$ indicates that the summations are to be carried over all $m$'s that correspond to open diffraction channels. In particular, for square integrability to hold, each summand in the equations above must be zero. It follows that for each open channel $m_{\sz}$ the following system holds,
\begin{equation}
\begin{cases}
\displaystyle{e^{i h k_{z,m_{\sz}}}\cph+e^{-i (a(\kx+2\pi m_{\sz})+h k_{z,m_{\sz}})}\cpa }=0\\
\displaystyle{e^{-i h k_{z,m_{\sz}}}\cph+e^{-i (a(\kx+2\pi m_{\sz})-h k_{z,m_{\sz}})}\cpa }=0
\end{cases}
\label{eq:72}
\end{equation}
For nontrivial solutions, the determinant of the above system must be zero. Thus, $\sin(2h k_{z,m_{\sz}})=0$ for each open channel $m_{\sz}$. Moreover, by considering the ratio of the field $\cpa$ to the field $\cph$ for all the open channels
it is deduced from system (\ref{eq:72}) that,
\begin{equation}
\cos(2h\kz)=\cos(2h\kzn)e^{-2\pi i a}=\cos(2h k_{z,1})e^{2\pi i a}=...
\label{eq:73}
\end{equation}
In particular, since for each open channel $m_{\sz}$ we have $\sin(2h k_{z,m_{\sz}})=0$, then $\cos(2h k_{z,m_{\sz}})=\pm 1$. Thus $e^{2\pi i a }=\pm 1$, hence $a=0$ or $a=\sh$.

The first two equations of system (\ref{eq:69}) determine $k$ and $h$ as functions of $\kx$ while the last equations determine the values of $\kx$. Thus as pointed out above, the values of $\kx$ at which bound states may exist become more sparse as the number of open diffraction channels increases.

In the case of two open diffraction channels, it is remarkable that these values of $\kx$ are dense in $[0,\pi]$. A proof of this statement is given in Section~\ref{sec:10}. It is shown there that the values of $\kx$ allowing for the existence of bound states occur in a double sequence $k^{n,l}_{x}$ where $n,l$ are positive integers. In the leading order of $R^2(\eec-1)$, the elements of the subsequence $k^{2n+1,l}_{x}$ are shown to have the form :
\begin{equation}
k^{2n+1,l}_{x}\approx\frac{\pi}{2r^2-1}+\frac{\pi^5(r^2-1)(4r^2-1)^4}{4(2r^2-1)^5}R^4(\eec-1)^2, \quad r=\frac{l}{2n+1}
\label{eq:56.1}
\end{equation}
The elements of the subsequence $k^{2n,l}_{x}$ are harder to derive due to a more intricate dependence on the system parameters. For the subsequence $k^{2n+1,l}_{x}$, the corresponding wavenumbers and distances between the arrays at which the bound states occur are proved to be obtained by substituting $k^{2n+1,l}_{x}$ into the following expressions:
\begin{equation}
k_{2n+1}(\kx)\approx 2\pi+\kx-\frac{8\pi^2\dlot(2\pi+\kx)}{2\pi+\kx}\,, \qquad h_{2n+1}(\kx)\approx\frac{(2n+1)\pi}{2\sqrt{2\pi \kx}}\left(1+\frac{\pi \dlot(2\pi+\kx)}{\kx}\right)
\label{eq:53.1}
\end{equation}
As in the case of bound states in the continuum I, bound states in the continuum II are shown to only occur under the following restriction on the the radius and dielectric constant of the scattering cylinders :
\begin{equation}
R\sqrt{\eec-1}<\frac{C\sqrt[4]{\kx}}{\sqrt{\pi-\kx}},\quad C\approx 2.016
\label{eq:51}
\end{equation}
In particular, when $\kx$ is close to $0$, the quantity $R\sqrt{\eec-1}$ must be small enough in order for bound states to exist at all. Figure~\ref{fig:2}(c) shows the regions in which Eq.(\ref{eq:51}) is valid.

The near field amplification observed in the case of one open diffraction channel persists when two channels are open. This can be established via an analysis similar to that of Section~\ref{sec:7}. Figure~\ref{fig:4}(c) gives an example of such an amplification.

When three or more diffraction channels are open; the equations $\sin(2hk_{z,m_{\sz}})=0$ for each open channel $m_{\sz}$, determine the parameters $\kx,\: k$ and $h$. In fact, if $n_{0}$, $n_{1}$, $n_{2},...$ are positive integers such that $2h\kz=n_{0}\pi$, $2h\kzn=n_{1}\pi$, $2h k_{z,1}=n_{2}\pi...$ then $2n^{2}_{0}\neq n^{2}_{1}+n^{2}_{2}$ and,
\begin{equation}
\kx=\frac{n^{2}_{1}-n^{2}_{2}}{2n^{2}_{0}-n^{2}_{1}-n^{2}_{2}}\, \pi\,,\quad h=\frac{1}{4\sqrt{2}}\sqrt{2n^{2}_{0}-n^{2}_{1}-n^{2}_{2}}\,,\quad k=\frac{\sqrt{(n^{2}_{1}+n^{2}_{2}-4n^{2}_{0})^{2}+4n^{2}_{1}n^{2}_{2}}}{2n^{2}_{0}-n^{2}_{1}-n^{2}_{2}}\, \pi
\label{eq:74}
\end{equation}
When only three channels are open, the integers $n_{0}$, $n_{1}$ and $n_{2}$ are only required to be in the order $n_{0}>n_{1}\geq n_{2}$. When four channels are open, the additional equation $\sin(2h k_{z,-2})=0$ in system (\ref{eq:69}) requires that the aforementioned integers satisfy the system,
\begin{equation*}
\begin{cases}
3n^{2}_{1}+n^{2}_{2}=3n^{2}_{0}+n^{2}_{3}\\
n_{0}\geq n_{1}>n_{2}\geq n_{3}
\end{cases}
\label{eq:75}
\end{equation*}
As more diffraction channels become available, there are more and more constraints on the integers $n_{i},\: i=0,1,2...$ Provided integers satisfying those constraints can be found, bound states will then be formed for double arrays for which the radius $R$ and the dielectric constant $\eec$ satisfy $\Delta(a,h,\kx,k)=0$ with $h$, $k$ and $\kx$ given by Eqs.(\ref{eq:74}) and $a\in \{0,\sh\}$. From Eqs.(\ref{eq:10}) and (\ref{eq:13}), it follows that $R$ and $\eec$ must be on curves,
\begin{equation*}
\frac{2}{k^2R^2(\eec-1)}+\ln(2\pi R)=C(n_{0},n_{1},...)
\label{eq:76}
\end{equation*}
where $k$ is given in Eq.(\ref{eq:74}) and $C$ is some constant that depends on the integers $n_{i},\: i=0,1,2,...$

\subsection{Bound states in the radiation continuum II: Two open  diffraction channels}
\label{sec:10}

Suppose that both the $0^{\tx{th}}$ and $-1^{\tx{st}}$ diffraction channels are open ,i.e., $\kx\in (0,\pi]$ and $2\pi-\kx<k<2\pi+\kx$. Conditions (\ref{eq:69}) then translate to the following system of equations:
\begin{equation}
\begin{cases}
\displaystyle{\frac{2\left(1-\cos(2\pi a)\cos(2h\kz)\cos(2h\kzn)\right)}{\kz\kzn}=\Psi^{+}\Psi^{-}}\vspace{0.2 cm}\\
\displaystyle{\Psi^{+}\left(\frac{1-\cos(2h\kz)}{\kz}+\frac{1-\cos(2\pi a)\cos(2h\kzn)}{\kzn}\right)+\Psi^{-}\left(\frac{1+\cos(2h\kz)}{\kz}+\frac{1+\cos(2\pi a)\cos(2h\kzn)}{\kzn}\right)=0}
\end{cases}
\label{eq:40}
\end{equation}
where $\Psi^{\pm}=\php\mp\phc$ for the functions $\php$ and $\phc$ of Eqs.(\ref{eq:14}). Also, recall that $a$ is necessarily $0$ or $\sh$ as derived from Eqs.(\ref{eq:73}).

The first equation of the system implies that $\Psi^{+}\Psi^{-}\geq 0$. If this inequality were to be strict, the second equation would not have held, and, therefore, $\Psi^{+}\Psi^{-}=0$. Thus either $\Psi^{+}=0$ or $\Psi^{-}=0$. Note that the functions $\Psi^{+}$ and $\Psi^{-}$ cannot vanish simultaneously. This can be verified by observing that,
\begin{equation*}
{\Psi^{+}}^2+{\Psi^{-}}^{2}=2\left(\Phi_{\ast}^{2}+\Phi_{c}^{2}\right)
\end{equation*}
and therefore, if the functions $\Psi^{+}$ and $\Psi^{-}$ were to vanish simultaneously; it would follow that $\phc=0$. But,
\begin{equation*}
\phc=\begin{cases}
     \displaystyle{\sum_{m\neq 0,-1}\frac{e^{-2h\qzm}}{\qzm}>0} &\text{if $a=0$}\\
     \displaystyle{\sum_{m\neq 0,-1}(-1)^{m}\frac{e^{-2h\qzm}}{\qzm}<0} &\text{if $a=\frac{1}{2}$ and $\kx\neq \pi$}\vspace{0.5 mm}\\
     \qquad 0 &\text{if $a=\frac{1}{2}$ and $\kx=\pi$}
     \end{cases}
     \end{equation*}
Thus $(a,\kx)=(\sh,\pi)$. But then $\kz=\kzn$ and the first equation in system (\ref{eq:40}) reads,
\begin{equation*}
1+\cos^{2}(2h\kz)=0
\end{equation*}
This is impossible and therefore at a bound state $\Psi^{+}$ and $\Psi^{-}$ do not vanish simultaneously. In particular, there are no bound states corresponding to the pair $(a,\kx)=(\sh,\pi)$. This is the reason this point was removed from the third level of Fig.~\ref{fig:2}(a).

For the remainder of the discussion it is assumed that $(a,\kx)\neq(\sh,\pi)$. System (\ref{eq:40}) then splits into two systems, namely,
\begin{equation}
\begin{cases}
\cos(2h\kz)=-1\\
\cos(2\pi a)\cos(2h\kzn)=-1\\
\Psi^{+}=0
\end{cases}
\qquad
\begin{cases}
\cos(2h\kz)=1\\
\cos(2\pi a)\cos(2h\kzn)=1\\
\Psi^{-}=0
\end{cases}
\label{eq:45}
\end{equation}
Thus to each value of $a$ corresponds a pair of systems whose solutions, if any, give rise to bound states in the continuum. For $a=0$, these systems are,
\begin{subequations}\label{eq:46}
\begin{equation}\label{eq:46.1}
\text{(A):}\begin{cases}
\cos(2h\kz)=-1\\
\cos(2h\kzn)=-1\\
\Psi^{+}=0
\end{cases}
\qquad
\text{(B):}\begin{cases}
\cos(2h\kz)=1\\
\cos(2h\kzn)=1\\
\Psi^{-}=0
\end{cases}
\end{equation}
while for $a=\sh$ they are,
\begin{equation}\label{eq:46.2}
\text{(C):}\begin{cases}
\cos(2h\kz)=-1\\
\cos(2h\kzn)=1\\
\Psi^{+}=0
\end{cases}
\qquad\quad
\text{(D):}\begin{cases}
\cos(2h\kz)=1\\
\cos(2h\kzn)=-1\\
\Psi^{-}=0
\end{cases}
\end{equation}
\end{subequations}
The existence of solutions to the last two equations in each system is proved first. Then the first equation of each system is added to show the existence of bound states.

For each positive integer $n$, the function $\Psi_{n}$ is defined by,
\begin{equation}
\begin{split}
 \Psi_{n}(k,\kx,a)&=\begin{cases}
                    \displaystyle{\Psi^{+}(a,\frac{n\pi}{2\kzn},k,\kx)} \quad \text{if $n$ is odd and $a=0$ or $n$ is even and $a=\frac{1}{2}$}\vspace{1 mm}\\
                    \displaystyle{\Psi^{-}(a,\frac{n\pi}{2\kzn},k,\kx)} \quad \text{if $n$ is even and $a=0$ or $n$ is odd and $a=\frac{1}{2}$}
                    \end{cases}\\
                  &=\frac{1}{2\pi\dlo(k)}+\sum_{m\neq 0,-1}\left(\frac{1}{2\pi(|m|+1)}-\frac{1-(-1)^{n+2a(m+1)}e^{-n\pi\qzm k^{-1}_{z,-1}}}{\qzm}\right)+\frac{1}{\pi}\left(\frac{3}{4}+\ln(2\pi R)\right)
                  \end{split}
                  \label{eq:47}
                  \end{equation}
where $k\in (2\pi-\kx,2\pi+\kx)$. Then the systems formed by the last two equations of each of systems (\ref{eq:46}) split into the countable set of systems
\begin{equation}
\begin{cases}
\displaystyle{h=\frac{n\pi}{2\sqrt{k^2-(2\pi-\kx)^2}}}\\
\Psi_{n}(k,\kx,a)=0
\end{cases}
\qquad n=1,2,3...
\label{eq:48}
\end{equation}

It is shown in Appendix~\ref{asec:3} that for each positive integer $n$, the function $k\mapsto \Psi_{n}$ is monotone decreasing on its domain $(2\pi-\kx,2\pi+\kx)$. It follows that if system (\ref{eq:48}) has a solution, then this solution is unique. Moreover, such a solution will only exist if and only if the limit of $k\mapsto\Psi_{n}$ at $2\pi-\kx$ is positive while the limit of $\Psi_{n}$ at $2\pi+\kx$ is negative. The first limit is,
\begin{equation}
\lim_{k\rightarrow (2\pi-\kx)^{+}} \Psi_{n}=\frac{1}{2\pi \dlo(2\pi-\kx)}+\sum_{m\neq 0,-1} \left(\frac{1}{2\pi(|m|+1)}-\frac{1}{\sqrt{(2\pi m+\kx)^2-(2\pi-\kx)^2}}\right)+\frac{1}{\pi}\left(\frac{3}{4}+\ln(2\pi R)\right)
\label{eq:49}
\end{equation}
As in the case of bound states in the continuum I, the requirement for this limit to be positive puts a restriction on the values of $R$ and $\eec$ that allow for the existence of bound states in the continuum II. An easily analyzable condition on these parameters is obtained by following the same procedure as in Section~\ref{sec:4}. First, the positivity condition is rewritten as,
\begin{align}
\frac{2}{\pi R^2(\eec-1)(2\pi-\kx)^2}>\sum_{m=1}^{\infty}\Bigg(\frac{1}{\sqrt{(2\pi m+\kx)^2-(2\pi-\kx)^2}}+\frac{1}{\sqrt{(2\pi(m+1)-\kx)^2-(2\pi-\kx)^2}}-&\frac{1}{\pi\sqrt{m(m+1)}}\Bigg)\nonumber\\
+\frac{1}{\pi}\Big(s-\frac{3}{4}-&\ln(2\pi R)\Big)
\label{eq:50}
\end{align}
where $$s=\sum_{m=1}^{\infty}\left(\frac{1}{\sqrt{m(m+1)}}-\frac{1}{2}\left(\frac{1}{m+1}+\frac{1}{m+2}\right)\right)\approx 0.691$$
The rearrangement of the series is made to ensure that all the summands in the series of Eq.(\ref{eq:50}) are nonnegative; they vanish at $\kx=\pi$. Also since the cylinders are thin,
it may be assumed that $R< \frac{1}{2\pi}e^{s-\frac{3}{4}}\approx 0.150$ so that $s-\frac{3}{4}-\ln(2\pi R)>0$. Thus all summands in Eq.(\ref{eq:50}) are nonnegative and hence the left hand side must be larger than each individual summand on the right hand side. To estimate the threshold value of $R\sqrt{\eec-1}$ below which bound states may exist, the first term $(m=1)$ is retained in the sum (\ref{eq:50}). This term is then written in the form $(\pi-\kx)\sqrt{\kx}g(\kx)$ to isolate its branch point at $\kx=0$ and its simple root at $\kx=\pi$. The minimization of $g(\kx)$ on $[0,\pi]$ produces the estimate (\ref{eq:51}) where $C$ is exactly,
\begin{equation*}
C=2^{\frac{5}{4}}\pi^{-\frac{3}{4}}\left(\min_{0\leq t\leq 1}\frac{(2-t)^2}{\sqrt{3-t}}\left(\frac{\sqrt{3-t}}{1+\sqrt{t}}-\frac{\sqrt{t}}{\sqrt{2}+\sqrt{3-t}}\right)\right)^{-\frac{1}{2}}\approx 2.016
\end{equation*}

As mentioned above, in addition to the requirement that the limit at $2\pi-\kx$ of $k\mapsto\Psi_{n}$ be positive, one must also require that the limit at $2\pi+\kx$ be negative in order for system (\ref{eq:48}) to have a solution. The latter limit is,
\begin{equation*}
\lim_{k\rightarrow (2\pi+\kx)^{-}}\Psi_{n}=\begin{cases}
                                    -\infty &\text{if $n$ is odd}\vspace{1.5 mm}\\
                                    \displaystyle{-\frac{n\pi}{\sqrt{2\pi\kx}}+\bigo(1)} &\text{if $n$ is even and $a=0$}\vspace{1 mm}\\
                                    \displaystyle{-\frac{1}{\sqrt{3\pi (\pi-\kx)}}+\bigo(1)} &\text{if $n$ is even and $a=\frac{1}{2}$}
                                    \end{cases}
                                    \label{eq:53}
                                    \end{equation*}
In particular, the limit is negative if $n$ is odd. If $n$ is even and $a=0$, the negativity condition may be ensured by choosing $\kx$ sufficiently close to zero or by choosing $n$ sufficiently large. If $n$ is even and $a=\sh$, then $\kx$ must be very close to $\pi$ for the limit to be negative. For parameters $R$ and $\eec$ satisfying condition (\ref{eq:50}), the conditions of existence of the solutions $\big(k_{n}(\kx),h_{n}(\kx)\big)$ to system (\ref{eq:48})
are summarized in Table~\ref{tb:1}. In the leading order of $\dlo(2\pi+\kx)$, approximate values of $k_{2n+1}$ and $h_{2n+1}$ are given by Eq.(\ref{eq:53.1})(See Appendix~\ref{asec:4} for derivation).

The approximate values for the wavenumbers $k_{2n}(\kx)$ are more difficult to find due to the dependence of their existence on the physical parameters of the system.
\begin{table}[!h]
\begin{center}
\begin{tabular}{|c|c|c|}
    \hline
        & $a=0$ & $a=\frac{1}{2}$ \\
    \hline
$n$ odd & $(k_{n},h_{n})$ exists $\forall \kx\in (0,\pi]$ & $(k_{n},h_{n})$ exists $\forall \kx\in (0,\pi)$\\
    \hline
$n$ even & $(k_{n},h_{n})$ exists for $n$ large or $\kx$ small & $(k_{n},h_{n})$ exists only for $\kx$ very close to $\pi$ \\
    \hline
    \end{tabular}
    \caption{Existence of solutions to systems (\ref{eq:48}). In particular, systems (A) and (D) in (\ref{eq:46}) always have solutions whereas (B) and (C) might not.}
    \label{tb:1}
\end{center}
\end{table}\\
\indent Since the conditions of existence of solutions to system (\ref{eq:48}) are now established,
the existence of solutions to systems (\ref{eq:46})
can be investigated. So far only  the last two equations in each of the latter systems have been used to determine  the values $k_{n}(\kx)$ and $h_{n}(\kx)$
for each $\kx\in (0,\pi]$ that are
susceptible to permit the existence of bound states. It follows that the first equations in each of systems (\ref{eq:46}) determine the values of $\kx$ that allow for the existence of bound states. In the coming paragraphs the set of these values is shown to be discrete and dense in $[0,\pi]$.

Let $n$ be a positive integer for which $k_{n}(\kx)$ exists for all $\kx\in(0,\pi]$ ($\kx\in (0,\pi)$ if $a=\sh$). Consider the function $\varphi_{n}$ defined by,
\begin{equation*}
\varphi_{n}(\kx)=2h_{n}(\kx)\sqrt{k^{2}_{n}(\kx)-k^{2}_{x}}=n\pi\sqrt{\frac{k^{2}_{n}(\kx)-k^{2}_{x}}{k^{2}_{n}(\kx)-(2\pi-k_{x})^2}}\,, \quad \kx\in (0,\pi)
\label{eq:54}
\end{equation*}
where the value $\kx=\pi$ is purposely left out and will be discussed later.

Collectively, the first equations of systems (\ref{eq:46}) are $\cos\left(\varphi_{n}(\kx)\right)=\pm 1$ and therefore, they have solutions if the range of $\varphi_{n}$ can be shown to contain even and odd integer multiples of $\pi$. That this is indeed the case follows from the continuity of $\varphi_{n}$ and its limits at $0$ and $\pi$. These are,
\begin{equation*}
\lim_{\kx\rightarrow 0^{+}} \varphi_{n}(\kx)=\infty \quad \text{and} \quad \lim_{\kx\rightarrow \pi^{-}}\varphi_{n}(\kx)=n\pi
\label{eq:56}
\end{equation*}
so that the range of $\varphi_{n}$ contains the interval $(n\pi,\infty)$. In particular, for each positive integer $l>n$, there exists a point $k^{n,l}_{x}\in (0,\pi)$ such that $\varphi_{n}(k^{n,l}_{x})=l\pi$ and therefore $\cos\left(\varphi_{n}(k^{n,l}_{x})\right)=(-1)^{l}$. This establishes the existence of bound states in the continuum II. As claimed, they exist for a discrete set of $\kx$ values in $(0,\pi)$, namely, the points $k^{n,l}_{x}$. To each of these points corresponds a specific wavenumber $k_{n,l}=k_{n}(k^{n,l}_{x})$ and a specific distance $h_{n,l}=h_{n}(k^{n,l}_{x})$ at which a bound state in the continuum exists. Note that the points $k^{n,l}_{x}$ depend on $R$, $\eec$ and the shift $a$ as is illustrated for instance for the points $k^{2n+1,l}_{x}$ in Eq.(\ref{eq:56.1}) (See Appendix~\ref{asec:4} for details). Note also that by definition, the values $k^{2n+1,2l+1}_{x}$ are solutions to system (A) of Eqs.(\ref{eq:46}) while the values $k^{2n+1,2l}_{x}$ are solutions to system (B) of the same set of systems and that the two sets of points do not overlap.

As far as the case $\kx=\pi$ is concerned, it was established in the beginning of this section that there can be no bound states if $a=\sh$. However, they do exist if $a=0$. This is because $\kz=\kzn$ at $\kx=\pi$ so that the first equations in systems (\ref{eq:46.1}) are superfluous and hence the existence of solutions to systems (\ref{eq:48}) alone suffices to guarantee the existence of bound states in the continuum II (Note that systems (\ref{eq:46.2}) become inconsistent as expected). Thus if $a=0$, the list of points $\{k^{n,l}_{x},\: l>n\}$ is to be completed by adding to it the point $\kx=\pi$. Keeping with the notation, this point is $k^{n,n}_{x}$ for each $n$ since $\varphi_{n}$ can be extended to $\pi$ by defining $\varphi_{n}(\pi)=n\pi$ so that $\cos(\varphi_{n}(\pi))=(-1)^n$. Thus the point $\kx=\pi$ is of infinite multiplicity in the list $\{k^{n,l}_{x}, l\geq n\}$ and hence is associated with an infinite set of bound states. Indeed for each positive integer $n$, sufficiently large if it is even, there exists a bound state at the wavenumber $k_{n,n}=k_{n}(\pi)$ and at the distance $2h_{n,n}=2h_{n}(\pi)$ between the two arrays of cylinders. All the other points $k^{n,l}_{x}$ change with the physical parameters of the cylinders and the indices $n,l$ and therefore none of them is certain to be repeated.

\indent It is remarkable that the set of points $k^{n,l}_{x}$ is dense in $[0,\pi]$ as is suggested by Eq.(\ref{eq:56.1}). To demonstrate this fact, consider an arbitrary interval $\text{I}_{\alpha\beta}=(\alpha,\beta)\subset (0,\pi)$. It will
be shown shortly that,
\begin{equation}
\lim_{n\rightarrow\infty} \Big(\varphi_{2n+1}(\alpha)-\varphi_{2n+1}(\beta)\Big)=\infty
\label{eq:57}
\end{equation}
Therefore, however small the interval $\text{I}_{\alpha\beta}$ may be, the interval $\varphi_{2n+1}(\text{I}_{\alpha\beta})$ contains positive integer multiples of $\pi$ for sufficiently large $n$. If $l\pi$ is such a multiple, then $k^{2n+1,l}_{x}\in \text{I}_{\alpha\beta}$. Thus the points $\{k^{2n+1,l}_{x}\}$ are dense in $(0,\pi]$.\\
\indent Before  establishing the limit (\ref{eq:57}), recall that $k_{n}=k_{n}(\kx)$ designates the solution to the equation $\Psi_{n}(k_{n}(\kx),\kx,a)=0$. The first task is to show that the sequence $\{ k_{2n+1}(\kx)\}^{\infty}_{n=1}$ converges for each fixed $\kx$.  To this end, let $\Psi_{\infty}$ be the function defined by,
\begin{equation}
\Psi_{\infty}(k,\kx)=\frac{1}{2\pi\dlo(k)}+\sum_{m\neq 0,-1} \left(\frac{1}{2\pi(|m|+1)}-\frac{1}{\qzm}\right)+\frac{1}{\pi}\left(\frac{3}{4}+\ln(2\pi R)\right)\,, \quad k\in(2\pi-\kx,2\pi+\kx)
\label{eq:58}
\end{equation}
For fixed $(a,\kx)$, the function $k\mapsto \Psi_{\infty}$ is the pointwise limit of the sequence of functions $k\mapsto \Psi_{n}$; it is also continuous and monotone decreasing on its domain $(2\pi-\kx, 2\pi+\kx)$ as is shown in Appendix~\ref{asec:3}. As $k\rightarrow (2\pi+\kx)^{-}$, $\Psi_{\infty}\rightarrow -\infty$ while the limit of $\Psi_{\infty}$ at $(2\pi-\kx)^{+}$ is exactly limit (\ref{eq:49}). In particular, the latter limit is positive as required by condition (\ref{eq:50}). It follows that for each $\kx \in (0,\pi]$, there exists a unique point $k_{\infty}(\kx)\in (2\pi-\kx,2\pi+\kx)$ such that $\Psi_{\infty}(k_{\infty}(\kx),\kx)=0$.\\
\indent Now for fixed $\kx$,
\begin{equation}
\Psi_{\infty}(k_{2n+1})=\Psi_{\infty}(k_{2n+1})-\Psi_{2n+1}(k_{2n+1})=
                                                                \begin{cases}
                                                                \displaystyle{\sum_{m=1}^{\infty}\left(\frac{e^{-(2n+1)\pi \qzm k^{-1}_{z,-1}}}{\qzm}+\frac{e^{-(2n+1)\pi q_{z,-m-1} k^{-1}_{z,-1}}}{q_{z,-m-1}}\right)}  &\text{if $a=0$}\vspace{2 mm}\\
                                                                \displaystyle{\sum_{m=1}^{\infty}\left(\frac{e^{-(2n+1)\pi \qzm k^{-1}_{z,-1}}}{\qzm}-\frac{e^{-(2n+1)\pi q_{z,-m-1} k^{-1}_{z,-1}}}{q_{z,-m-1}}\right)(-1)^{m+1}} &\text{if $a=\frac{1}{2}$}
                                                                \end{cases}
                                                                \label{eq:59}
                                                                \end{equation}
where it is understood that in the two series on the right, the terms $\qzm$ and $\kzn$ are to be evaluated at $k=k_{2n+1}$. The sum of the first series is obviously positive. The sum of the second series is also nonnegative as it is the sum of an alternating series whose terms decrease in absolute value. Thus $\Psi_{\infty}(k_{2n+1})\geq 0$.\\
\indent As $k\mapsto \Psi_{\infty}$ is a decreasing function it follows that,
\begin{equation}
k_{2n+1}(\kx)\leq k_{\infty}(\kx)<2\pi+\kx \,, \quad \forall n=1,2,3... \quad \text{and} \quad \forall \kx\in (0,\pi]
\label{eq:60}
\end{equation}
In particular, $q_{z,1}=\sqrt{(2\pi+\kx)^2-k^{2}_{2n+1}}\geq \sqrt{(2\pi+\kx)^2-k^{2}_{\infty}}>0$ and therefore $q_{z,1}$ does not converge to $0$ as $n\rightarrow \infty$. Hence,
\begin{equation*}
\Psi_{\infty}(k_{2n+1})= \bigo\left(\frac{e^{-(2n+1)\pi q_{z,1} k^{-1}_{z,-1}}}{q_{z,1}}\right) \xrightarrow[n\rightarrow\infty]{} 0=\Psi_{\infty}(k_{\infty})
\label{eq:61}
\end{equation*}
Since the function $k\mapsto\Psi_{\infty}$ is continuous and bijective for each fixed $\kx\in (0,\pi]$, it follows that $k_{2n+1}(\kx)\rightarrow k_{\infty}(\kx)$ as $n\rightarrow \infty$. Note that without condition (\ref{eq:60}), $q_{z,1}$ could converge to $0$ causing the sequence $\{ \Psi_{\infty}(k_{2n+1})\}^{\infty}_{n=1}$ to diverge as indicated by Eqs.(\ref{eq:59}). This is the reason the subsequence $\{ k_{2n} \}_{n=1}^{\infty}$ had to be excluded. For this subsequence, it can be shown through an analysis similar to the above that $k_{2n}(\kx)\geq k_{\infty}(\kx)\,, \forall n=1,2,3...$ and therefore more work would be needed to show that no subsequence of the sequence $\{k_{2n}\}_{n=1}^{\infty}$ converges to $2\pi+\kx$.\\
\indent From the convergence of the sequence $\{k_{2n+1}(\kx)\}_{n=1}^{\infty}$ it is deduced that,
\begin{equation}
\lim_{n\rightarrow\infty} \frac{1}{(2n+1)\pi}\Big(\varphi_{2n+1}(\alpha)-\varphi_{2n+1}(\beta)\Big)=\varphi_{\infty}(\alpha)-\varphi_{\infty}(\beta),\quad \varphi_{\infty}(\kx)= \sqrt{\frac{k^2_{\infty}(\kx)-k^{2}_{x}}{k^{2}_{\infty}(\kx) -(2\pi-\kx)^2}}
\label{eq:62}
\end{equation}
In Appendix~\ref{asec:3}  the function $\varphi_{\infty}$
is shown to be
strictly decreasing so that $\varphi_{\infty}(\alpha)-\varphi_{\infty}(\beta)>0$. This establishes limit (\ref{eq:57}) and thereby the density of the points $\{k^{2n+1,l}_{x} \}$ in the interval $(0,\pi]$. Moreover, if $|A_{n}(\alpha,\beta)|$ is the cardinality of the set $A_{n}(\alpha,\beta)=\{l : k^{n,l}_{x}\in (\alpha,\beta) \}$ then,
\begin{equation*}
|A_{2n+1}(\alpha,\beta)|= (2n+1) \Big(\varphi_{\infty}(\alpha)-\varphi_{\infty}(\beta)\Big)+o(n)\xrightarrow[R^2(\eec-1)\rightarrow 0]{} \frac{(2n+1)}{\sqrt{2}}\left(\sqrt{1+\frac{\pi}{\alpha}}-\sqrt{1+\frac{\pi}{\beta}}\right)+o(n)
\label{eq:63}
\end{equation*}
The last limit follows from the fact that as $R^2(\eec-1)\rightarrow 0$ then $k_{\infty}(\kx)\rightarrow 2\pi+\kx$. \\
\indent Lastly, from Eqs.(\ref{eq:9}) and (\ref{eq:12}), the analytic expressions of the bound states $E_{n,l}$
can be obtained
for positive integers $n$ and $l$ such that $l\geq n$ if $a=0$ and $l>n$ if $a=\sh$. As mentioned in prior sections, each of these states can only be determined up to a multiplicative constant which is chosen to be the value $E_{n,l}(-h_{n,l}\ez)$ of the electric field on the cylinder at $(0,-h_{n,l})$. In terms of this value, the electric field on the cylinder at $(a,h_{n,l})$ is then,
\begin{equation*}
E_{n,l}(a\ex+h_{n,l}\ez)=(-1)^{n+2a+1}e^{i a k^{n,l}_{x}}E_{n,l}(-h_{n,l}\ez)
\label{eq:67}
\end{equation*}
and everywhere off the scattering cylinders it is,
\begin{equation}
E_{n,l}(\rr)=2\pi i\dlo(k_{n,l}) E_{n,l}(-h_{n,l}\ez)\sum_{m} \frac{e^{i x(k^{n,l}_{x}+2\pi m)}}{k^{n,l}_{z,m}}\left(e^{i|z+h_{n,l}|k^{n,l}_{z,m}}+(-1)^{n+2a(m+1)+1}e^{i|z-h_{n,l}|k^{n,l}_{z,m}}\right)
\label{eq:68}
\end{equation}
where $k^{n,l}_{z,m}=\sqrt{k^{2}_{n,l}-(k^{n,l}_{x}+2\pi m)^2}$.\\

  \appendix
  \begin{center}
  {\bf Appendix}
  \end{center}
  \section{The Lippmann-Schwinger integral equation}
  \label{asec:1}
  \indent In this section of the appendix, it is proved
that the solution to the Lippmann-Schwinger integral equation in Eq.(\ref{eq:6}) solves Eq.(\ref{eq:2}). This will be done by considering the functions involved as distributions acting on smooth functions of compact support. \\
  \indent To do so, only locally integrable
solutions to Eq.(\ref{eq:2}) are sought.
In this setting, all the functions involved in Eq.(\ref{eq:6}), namely; $\eik$, $\eo$, $\ee \eo$, $(\ee-1)\eo$, $G$ and $ \left( (\ee-1) \eo \right) \ast G$, are locally integrable.\\
  \indent Let then $\varphi$ be a smooth function of compact support and $\eo$ be a locally integrable solution to Eq.(\ref{eq:6}). Then
  \begin{equation}
  \langle \Delta \eo+k^2 \ee \eo, \varphi \rangle = k^2 \langle \ee \eo, \varphi \rangle-k^2 \langle E_{i}, \varphi \rangle+\frac{k^2}{4 \pi} \langle\left( (\ee-1)\eo \right) \ast G,\Delta \varphi \rangle
  \label{eq:a1}
  \end{equation}
  where $E_{i}(\rr)=\eik$. Since none of the distributions $(\ee-1)\eo$ and $G$ is compactly supported, the convolution used here is to be understood in the sense of the usual convolution of functions. Therefore, the last integral of Eq.(\ref{eq:a1}) may be interpreted as
  \begin{equation}
  \langle\left( (\ee-1)\eo \right) \ast G,\Delta \varphi \rangle=\int (\ee(\ro)-1)\eo(\ro) \langle G_{\ro},\Delta \varphi \rangle d\ro
  \label{eq:a2}
  \end{equation}
  where $G_{\ro}(\rr)=G(\rr|\ro)$ and satisfies the distributional Helmholtz equation
  \begin{equation}
  \Delta G_{\ro}+k^2 G_{\ro} = -4\pi \delta_{\ro}
  \label{eq:a3}
  \end{equation}
  Therefore
  \begin{equation*}
  \langle G_{\ro},\Delta \varphi \rangle=\langle \Delta G_{\ro},\varphi \rangle=-k^2\langle G_{\ro},\varphi \rangle-4\pi \langle \delta_{\ro},\varphi \rangle
  \end{equation*}
  It follows that
  \begin{equation}
  \langle \left( (\ee-1)\eo \right) \ast G,\Delta \varphi \rangle=-k^2 \langle \left( (\ee-1)\eo \right) \ast G,\varphi \rangle-4 \pi \langle (\ee-1)\eo,\varphi \rangle
  \label{eq:a5}
  \end{equation}
  Thus $\langle \Delta \eo+k^2 \ee \eo, \varphi \rangle=0$ for all test functions $\varphi$.\\
  \indent Note that the only difference between the proof that Eq.(\ref{eq:6}) solves Eq.(\ref{eq:2}) in the case of the finite array and that of an infinite array is the way Eq.(\ref{eq:a5}) is derived from Eq.(\ref{eq:a2}). Indeed, if the array of cylinders is finite then the convolution of Eq.(\ref{eq:a2}) is in the distributional sense as $(\ee-1)\eo$ would be a compactly supported distribution. Therefore one could establish Eq.(\ref{eq:a5}) immediately from Eq.(\ref{eq:a3}) and the identity of distributional convolution
  \begin{equation}
  \langle \left( (\ee-1)\eo \right) \ast G,\Delta \varphi \rangle=\langle \left( (\ee-1)\eo \right) \ast \Delta G, \varphi \rangle
  \label{eq:a6}
  \end{equation}
  This identity is not immediate in the case of the infinite array because, as mentioned above, none of the convoluted functions has compact support. In fact, the function $\left( (\ee-1) \eo \right) \ast G$ is a conditionally convergent series so that integration by parts cannot be applied to establish Eq.(\ref{eq:a6}).\\

  \section{Solution of the Lippmann-Schwinger integral equation: zero radius approximation}
  \label{asec:2}
  \indent Here an approximate solution to the Lippmann-Schwinger integral equation is established
in the small radius approximation. In the usual theory of scattering from small particles \cite{b19,b20,b21,b22}, the problem is solved with high accuracy by assuming that far from the scattering region the solution is a linear superposition of the waves scattered by each individual particle. Due to the cylindrical
geometry of the dielectric scatterers, it turns out that similar approximations may be made to solve the scattering problem on the double array but with solutions which are valid everywhere off the scatterers even in the region between the two grating structures.
The validity of solution in this extended region
can be used to find the so-called
 hot spots (see Fig.~\ref{fig:3}) where the magnitude
of the electromagnetic fields peaks.\\
  \indent The structure considered is shown in Fig.~\ref{fig:1}(a). The cylinders in the structure are labeled as $C_{m,n}$ where $n$ is either $1$ or $-1$ depending on whether the cylinder is on the right or left array. The integer $m$ refers to the x-coordinate of the cylinder's axis. In particular, for the right array cylinders, the axes are positioned at $\rr_{m,1}=(m+a)\ex+h\ez$ and those of the left array are at positions $\rr_{m,-1}=m\ex-h\ez$.\\
  \indent The Lippmann-Schwinger integral equation may then be written as a sum over all cylinders as
  \begin{equation}
  \eo(\rr)=\eik+\frac{ik^2(\eec-1)}{4}\sum_{m,n} \int_{C_{m,n}} \eo(\ro) H_{\sz}(k|\rr-\ro|) d\ro
  \label{eq:b1}
  \end{equation}
  Far from the scatterers, each of the integrals is well approximated through the mean value theorem by
  \begin{equation*}
  \int_{C_{m,n}} \eo(\ro)G(\rr |\ro) d\ro \approx i \pi^{2} R^{2} e^{ i m k_{x} } \hz (k |\rr-\rmn|) \eo(\ron)
  \end{equation*}
  so that the far field may be expressed in terms of the fields on the cylinders $C_{0,\pm 1}$ as
  \begin{equation}
  \eo(\rr)=\eik+i\pi\delta_{_{\sz}}(k)\sum_{n=\pm 1}\eo(\ron) \sum_{m=-\infty}^{\infty} e^{i m \kx} H_{\sz}(k|\rr-\rmn|) \quad \text{for} \quad \delta_{_{\sz}}(k)=\frac{1}{4}k^2R^2(\eec-1)
  \label{eq:b3}
  \end{equation}
  The claim is that this approximation remains valid in the near region too. This may be established by means of the Bessel function expansions of the field inside the cylinders and the Hankel function $H_{\sz}$. To this end, let $\ro$ be a position vector on the cylinder $C_{m,n}$, then $\ro=\rmn+\uu$ with $u=|\uu | \leq R$. If $\rr$ is a position vector off the scatterers, then $|\rr-\rmn|>R$ and therefore
  \begin{equation*}
  H_{\sz}(k|\rr-\ro|)=\sum_{\nu=-\infty}^{\infty} e^{i\nu(\theta_{m,n}-\theta)}J_{\nu}(ku)H_{\nu}(k|\rr-\rmn|)
  \end{equation*}
  where $\theta$ and $\theta_{m,n}$ are the angles between the x-axis and the vectors $\uu$ and $\rr-\rmn$ respectively. On the other hand, the field inside the cylinder $C_{m,n}$ is given by
  \begin{equation}
  \eo(\ro)=e^{i m \kx} \sum_{\nu=-\infty}^{\infty} \alpha_{\nu,n} e^{i \nu \theta} J_{\nu}(k'u)
  \label{eq:b5}
  \end{equation}
  with $k'=n_{c}k$ for the index of refraction $n_{c}$ of the cylinders and the coefficients $\alpha_{\nu,n}$ given by
  \begin{equation*}
  \alpha_{\nu,n}=\frac{1}{2 \pi J_{\nu}(k'R)} \int_{0}^{2\pi} e^{-i \nu \theta} \eo(R {\bf e}_{r}+\ron) d\theta
  \end{equation*}
  Here ${\bf e}_{r}=\ex \cos \theta+ \ez \sin \theta$ is the usual radial vector of polar coordinates. In particular, $\alpha_{\nu,n}$ is at most of the order of $(kR)^{-|\nu|}$ in the limit of thin cylinders.

  It follows that
  \begin{equation*}
  \int_{C_{m,n}} \eo(\ro)G(\rr |\ro) d\ro= 2 \pi^{2} i e^{i m k_{x}} \sum_{\nu=-\infty}^{\infty} \alpha_{\nu,n} e^{i \nu \theta_{m,n}} H_{\nu}(k|\rr-\rmn|) \int_{0}^{R} J_{\nu}(k'u)J_{\nu}(k u) u du
  \end{equation*}
  \indent As the integral in the above series is of the order of $(kR)^{2|\nu|+2}$, it is then justified to approximate the series by its $0^{th}$ summand in the limit $kR\ll 1$, so that
  \begin{equation}
  \int_{C_{m,n}} \eo(\ro)G(\rr |\ro) d\ro\approx i \pi^2 R^2 e^{i m k_{x}} \alpha_{\sz,n} \hz (k|\rr-\rmn|)
  \label{eq:b4.1}
  \end{equation}
  \indent The value of $\alpha_{\sz,n}$ may then be recovered by setting $\uu=\mathbf{0}$ in Eq.(\ref{eq:b5}). It is $\eo(\ron)$. This establishes Eq.(\ref{eq:b3}) everywhere off the scatterers. In particular, the fields are determined by the knowledge of their values $\eo(\rr_{\sz,\pm 1})$ on the cylinders $C_{\sz,\pm 1}$ alone.

  To determine the values $\eo(\rr_{\sz,\pm 1})$, let $n$ be either $1$ or $-1$ and, $\rr_{\sz,n}$ be substituted for $\rr$ in Eq.(\ref{eq:b1}). The latter equation becomes,
  \begin{equation}
  \begin{split}
  \eo(\rr_{\sz,n})=e^{i\kk\cdot\rr_{\sz,n}}+\frac{i k^2(\eec-1)}{4}\bigg(\int_{C_{0,n}}\eo(\ro)H_{\sz}(k|\rr_{\sz,n}-\ro|)d\ro&+\sum_{m\neq 0} \int_{C_{m,n}} \eo(\ro)H_{\sz}(k|\rr_{\sz,n}-\ro|)d\ro\\
   &+\sum_{m} \int_{C_{m,-n}} \eo(\ro)H_{\sz}(k|\rr_{\sz,n}-\ro|)d\ro\bigg)
   \end{split}
   \label{eq:b6.1}
   \end{equation}
  where the integral over $C_{0,n}$ has been isolated due to the singularity of its integrand at $\ro=\rr_{\sz,n}$. To approximate this particular integral, Eq.(\ref{eq:b5}) is used to obtain in the leading order of $kR$;
  \begin{equation*}
  \int_{C_{0,n}} \eo(\ro)H_{\sz}(k|\rr_{\sz,n}-\ro|)d\ro=2\pi\eo(\rr_{\sz,n})\int_{0}^{R} J_{\sz}(k'u)H_{\sz}(ku)u du \approx \pi R^2\eo(\rr_{\sz,n})\left(1+\frac{2i}{\pi}\left(\gamma+\ln\left(\frac{kR}{2}\right)-\frac{1}{2}\right)\right)
  \end{equation*}
  where $\gamma$ is the Euler constant. All the other integrals in Eq.(\ref{eq:b6.1}) obey the estimate (\ref{eq:b4.1}). By taking $n$ successively equal to $1$ then to $-1$, the following system is obtained:
    \begin{equation}
          \begin{cases}
          \Phi_{\sz} \cpa + \Phi^{+} \cph =\displaystyle{\frac{i}{2\pi\delta_{_{\sz}}(k)} e^{ i (a k_{x} + h k_{z})}}\vspace{2 mm}\\
          \Phi^{-} \cpa + \Phi_{\sz} \cph =\displaystyle{\frac{i}{2\pi\delta_{_{\sz}}(k)} e^{- i h k_{z}}}
          \end{cases}
          \label{eq:b9}
  \end{equation}
 The functions $\Phi_{\sz}$,$\Phi^{+}$ and $\Phi^{-}$ are
 \begin{align*}
  &\Phi_{\sz}(k,\kx)=\frac{i}{2\pi\delta_{_{\sz}}(k)}+\frac{1}{2} \left( \sum_{m \neq 0} e^{i m k_{x} } \hz (k|m|)+1+\frac{2i}{\pi} \left(\gamma +\ln \left(\frac{ k R}{2} \right)-\frac{1}{2}\right) \right) \nonumber\\
  &\Phi^{\pm}(a,h,k,\kx)=\frac{1}{2} \sum_{m} e^{i m k_{x}} \hz (k|(m \mp a)\ex +h \ez|)
  \end{align*}
 The variants of these functions in Eqs.(\ref{eq:10}) are obtained through the formulas,
 \begin{subequations}\label{eq:b10}
 \begin{equation}\label{eq:b10.1}
 \frac{1}{2}\sum_{m=-\infty}^{\infty} e^{i m\kx} H_{\sz}(k|\rr-m\ex|)=\sum_{m=-\infty}^{\infty} \frac{e^{i(x(\kx+2\pi m)+|z|\kzm)}}{\kzm},\quad \rr\neq \mathbf{0}
 \end{equation}
 \begin{equation}\label{eq:b10.2}
 \frac{1}{2}\sum_{m\neq 0} e^{i m\kx}H_{\sz}(k|m|)=\sum_{m=-\infty}^{\infty}\left(\frac{1}{\kzm}-\frac{1}{2\pi(|m|+1)}\right)-\frac{1}{2}-\frac{i}{\pi} \left(\gamma+\ln\left(\frac{k}{4\pi}\right)-\frac{1}{2}\right)
 \end{equation}
 \end{subequations}
 Relation (\ref{eq:b10.1}) can be proved by substituting the plane wave representation of the Hankel function
 \begin{equation*}
  H_{\sz} (k r)=\frac{i}{\pi^2} \int\int \frac{e^{i (x K_{x}+z K_{z})}}{k^2-K^{2}_{x}-K^{2}_{z}+i \eta} dK_{z}dK_{x}, \quad \eta \rightarrow 0^{+}
  \end{equation*}
 into the left side of (\ref{eq:b10.1}) and carrying out the integration with respect to $K_{z}$ followed by an application of the Poisson summation formula to evaluate the integral with respect to $K_x$. Relation (\ref{eq:b10.2}) is the obtained from (\ref{eq:b10.1}) in the limit $\rr\rightarrow \mathbf{0}$, where for the term $m=0$ in the left side of Eq.(\ref{eq:b10.1}), the asymptotic expansion of the Hankel function for a small argument has to be used.

 When the determinant of the system (\ref{eq:b9}) is nonzero, Eq.(\ref{eq:2}) has a unique solution. Otherwise, the homogeneous Lippmann-Schwinger equation admits nonzero solutions: the bound states. These states can then be arbitrarily superposed to obtain the general solution. The various expressions for these bound states in Sections~\ref{sec:2} and \ref{sec:10} are obtained by applying formula (\ref{eq:b10.1}) to Eq.(\ref{eq:b3}) in the absence of the incident wave, i.e., by omitting the term $\eik$.\\

 \section{Bound states in the continuums I and II (Complements)}
 \label{asec:3}
 \indent \indent This section of the Appendix gives some of the technical details omitted in sections~\ref{sec:4} and~\ref{sec:10}. First,  Eq.(\ref{eq:27}) is stablished for the sequence $\{c_{m}\}_{m=1}^{\infty}$ defined in Eq.(\ref{eq:25}). In the case under consideration, the sequence $\{\qzm\}$ is in the order,
 \begin{equation*}
 q_{z,-1}\leq q_{z,1} <q_{z,-2}\leq q_{z,2}<...
 \label{eq:c1}
 \end{equation*}
 with equalities occurring when $\kx=0$. Since the function $f:t\in(0,\infty) \mapsto t^{-1}e^{-2h t}$ is strictly decreasing and $c_{m}=f(q_{z,-m})-f(q_{z,m})$, it follows that $c_{m}\geq 0$ with equality holding only if $\kx=0$.\\
 \indent Now suppose that $\kx\neq 0$ and, hence,  $c_{m}>0,\: m=1,2,3...$. To complete the proof of Eq.(\ref{eq:27}), it suffices to show that,
 \begin{equation}
 \frac{c_{m+1}}{c_{m}}\leq e^{-4\pi h} \quad \text{and} \quad \lim_{m\rightarrow\infty} \frac{c_{m+1}}{c_{m}}=e^{-4\pi h}
 \label{eq:c2}
 \end{equation}
 To establish the first of conditions (\ref{eq:c2}), the ratio of $c_{m+1}$ to $c_{m}$ is rewritten as,
 \begin{equation}
 \frac{c_{m+1}}{c_{m}}=\frac{\alpha_{-m-1}}{\alpha_{-m}}\frac{1-\frac{\alpha_{m+1}}{\alpha_{-m-1}}}{1-\frac{\alpha_{m}}{\alpha_{-m}}},\quad \alpha_{m}=\frac{e^{-2h\qzm}}{\qzm}
 \label{eq:c3}
 \end{equation}
 Next, the following chain of conclusions holds:
 \begin{equation*}
 \begin{cases}
 q_{z,m}+q_{z,-m-1}\geq q_{z,-m}+q_{z,m+1}\\
 q_{z,m}q_{z,-m-1}\geq q_{z,-m}q_{z,m+1}
 \end{cases}
 \Rightarrow \frac{\alpha_{-m}\alpha_{m+1}}{\alpha_{-m-1}\alpha_{m}}\geq 1 \Rightarrow \frac{1-\frac{\alpha_{m+1}}{\alpha_{-m-1}}}{1-\frac{\alpha_{m}}{\alpha_{-m}}}\leq 1 \Rightarrow \frac{c_{m+1}}{c_{m}}\leq \frac{\alpha_{-m-1}}{\alpha_{-m}}\leq e^{2h(q_{z,-m}-q_{z,-m-1})}
 \label{eq:c4}
 \end{equation*}
 The first of conditions (\ref{eq:c2}) then follows as $q_{z,-m}-q_{z,-m-1}\leq -2\pi$. The limit in (\ref{eq:c2}) follows from (\ref{eq:c3}) and the limits,
 \begin{equation*}
 \lim_{m\rightarrow\infty}\frac{\alpha_{-m-1}}{\alpha_{-m}}=e^{-4\pi h} \qquad \lim_{m\rightarrow \infty }\frac{\alpha_{m}}{\alpha_{-m}}=e^{-4h\kx}\neq 1
 \label{eq:c5}
 \end{equation*}
 \indent Second, the formula (\ref{eq:26}) is proved. This is done by a repetitive application of Abel's partial summation formula. Let $u_{n}$ be defined for each $n=0,1,2,...$ by,
 \begin{equation*}
 u_{n}=\sum_{m=1}^{\infty} c_{m+n}\sin(2\pi a m)
 \label{eq:c6}
 \end{equation*}
 The objective is to show that another expression of $-u_{0}$ is (\ref{eq:26}). By Abel's partial summation formula,
 \begin{equation*}
 \begin{split}
 u_{n}&=\sum_{m=1}^{\infty}\left(c_{m+n}-c_{m+n+1}\right)\frac{\sin(\pi a m)\sin(\pi a (m+1))}{\sin(\pi a)}\\
       &=\cot(\pi a)\sum_{m=1}^{\infty}\left(c_{m+n}-c_{m+n+1}\right)\sin^2(\pi a m)+\frac{1}{2}\sum_{m=1}^{\infty}\left(c_{m+n}-c_{m+n+1}\right)\sin(2\pi a m)\\
       &=\cot(\pi a)\sum_{m=1}^{\infty}\left(c_{m+n}-c_{m+n+1}\right)\sin^2(\pi a m)+\frac{1}{2}u_{n}-\frac{1}{2}u_{n+1}
       \end{split}
       \label{eq:c7}
       \end{equation*}
 Thus,
 \begin{equation*}
 u_{0}=(-1)^{N+1}u_{N+1}+2\cot(\pi a)\sum_{m=1}^{\infty}\left(c_{m}+2\sum_{n=1}^{N}(-1)^{n}c_{m+n}+(-1)^{N+1}c_{m+N+1}\right)\sin^2(\pi a m),\quad \forall N=1,2,3,...
 \label{eq:c8}
 \end{equation*}
 By using the first of conditions (\ref{eq:c2}) it is
straightforward
that $u_{N+1}\rightarrow 0$ as $N\rightarrow \infty$, and Eq.(\ref{eq:26}) follows.\\
 \indent Third, the functions $k\mapsto\Psi_{n}(k,\kx,a)$ defined in Eq.(\ref{eq:30}) are proved to be monotonically decreasing, i.e., $\partial_{k}\Psi_{n}<0$. This derivative
reads
 \begin{equation*}
 \frac{\partial \Psi_{n}}{\partial k}=-\frac{4}{\pi R^2 k^3 (\eec-1)}-\sum_{m\neq 0}\frac{k}{q^{3}_{z,m}}\left(1-(-1)^n\cos(2\pi a m) e^{-n\pi \qzm k^{-1}_{z}}\left(1+n\pi\left(\frac{\qzm}{\kz}+\frac{q^{3}_{z,m}}{k^{3}_{z}}\right)\right)\right)
 \label{eq:c9}
 \end{equation*}
 Now, if $t>0$ and $n$ is a positive integer; then $e^{-t}\left(1+t\left(1+(n\pi)^{-2}t^2\right)\right)\leq 1$. Setting $t=n\pi\qzm k^{-1}_{z}$ shows that all summands are positive and hence $\partial_{k}\Psi_{n}<0,\:\forall n=1,2,3,...$. The functions $k\mapsto\Psi_{n}(k,\kx)$ and $k\mapsto \Psi_{\infty}(k,\kx)$ defined in Eqs.(\ref{eq:47}) and (\ref{eq:58}) are shown to be decreasing in a similar fashion.\\
 \indent Lastly,  the function $\kx\mapsto \varphi_{\infty}$ defined in Eq.(\ref{eq:62}) is shown to be monotonically decreasing on $(0,\pi)$. To establish this fact, note that since $\partial_{k}\Psi_{\infty}<0$, the implicit function theorem implies that the function $\kx\mapsto k_{\infty}(\kx)$ is continuously differentiable and $k^{'}_{\infty}(\kx)=-\partial_{\kx}\Psi_{\infty}(\partial_{k} \Psi_{\infty})^{-1}$. Now,
\begin{equation*}
\frac{\partial \Psi_{\infty}}{\partial \kx}=\sum_{m=1}^{\infty}\left(\frac{2\pi m +\kx}{\sqrt{(2\pi m+\kx)^2-k^2}}-\frac{2\pi m+2\pi-\kx}{\sqrt{(2\pi m+2\pi-\kx)^2-k^2}}\right)>0
\label{eq:c10}
\end{equation*}
Hence $k^{'}_{\infty}(\kx)>0$. By logarithmic differentiation, it follows that,
\begin{equation*}
\frac{\varphi^{'}_{\infty}(\kx)}{\varphi_{\infty}(\kx)}=\frac{k_{\infty}k^{'}_{\infty}\left(k^{2}_{x}-(2\pi-\kx)^2\right)-\kx\left(k^{2}_{\infty}- (2\pi-\kx)^2\right)-(2\pi-\kx)\left(k^{2}_{\infty}-k^{2}_{x}\right)}{\left(k^{2}_{\infty}-k^{2}_{x}\right)\left(k^{2}_{\infty}-(2\pi-\kx)^2\right)}<0
\label{eq:c11}
\end{equation*}
since $0<\kx< 2\pi-\kx<k_{\infty}(\kx)$.

\section{Approximations}
\label{asec:4}
\indent In this section of the Appendix, it is outlined
how the approximations in Eqs.(\ref{eq:16.1}), (\ref{eq:37}), (\ref{eq:53.1}) and (\ref{eq:56.1}) can be  obtained. The computations of the wavenumbers in the first three of the latter equations being similar, so only Eqs.(\ref{eq:37})
is stablished, and the discussion is finished by proving Eq.(\ref{eq:56.1}). Suppose first that only one diffraction channel is open and $(a,\kx)$ is in the set $L$ of Eq.(\ref{eq:29}). The objective is to approximate the value $k_{n}\in (\kx,2\pi-\kx)$ such that $\Psi_{n}(k_{n},\kx,a)=0$ for the function $\Psi_{n}$ defined in Eq.(\ref{eq:30}). This is done by identifying the leading terms in $\Psi_{n}$ near the wavenumber $k_{n}$.

Since the dielectric cylinders forming the double array are assumed to be thin in comparison to the wavelength ,i.e., $k R\ll 1$, the quantity $(2\pi\dlo(k))^{-1}$ in the expression of $\Psi_{n}$ is large. Consequently, the wavenumber $k_{n}$ such that $\Psi_{n}(k_{n},\kx,a)=0$ must be close to the diffraction threshold $2\pi-\kx$ so that the term $q^{-1}_{z,-1}$ in $\Psi_{n}$ is large enough to compensate for the magnitude of $(2\pi\dlo(k))^{-1}$. Thus a first approximation for the wavenumber $k_{n}$ may be found by solving the equation,
\begin{equation*}
\frac{1}{2\pi\dlo(k)}-\frac{1-(-1)^{n}\cos(2\pi a) e^{-n\pi \qzn k^{-1}_{z}}}{\qzn}=0
\label{eq:d1}
\end{equation*}
which is obtained by keeping only the leading terms in the expression of $\Psi_{n}$ near $k_{n}$. When $(-1)^{n}\cos(2\pi a)=1$, then the equation becomes,
\begin{equation}
\frac{1}{2\pi\dlo(k)}-\frac{1-e^{-n\pi \qzn k^{-1}_{z}}}{\qzn}=0
\label{eq:d2}
\end{equation}
In particular if $n$ is not large, this equation has no roots since as $\qzn$ becomes smaller, then the second summand gets closer to $n\pi k^{-1}_{z}$ and hence is much smaller than the first summand. This was to be expected since in the case $(-1)^{n}\cos(2\pi a)=1$, it was already established that bound states exist only for sufficiently large $n$. Also, the initial integer $n$ at which the wavenumber $k_{n}$ exists for $(-1)^n\cos(2\pi a)=1$ grows as $k R\rightarrow 0$. This makes it impossible to provide a good approximation for the exponential term in Eq.(\ref{eq:d2}) that would allow a perturbative solution of the said equation. This complexity disappears when $(-1)^{n}\cos(2\pi a)\neq 1$. In this case and for $n$ not too large, the approximation $e^{-n\pi \qzn k^{-1}_{z}}\approx 1$ is valid and Eq.(\ref{eq:d2}) becomes,
\begin{equation*}
\frac{1}{2\pi \dlo(k)}-\frac{1-(-1)^n \cos(2\pi a)}{\qzn}=0
\label{eq:d3}
\end{equation*}
Thus,
\begin{equation}
k_{n}\approx \sqrt{(2\pi-\kx)^2-4\pi^2(1-(-1)^{n}\cos(2\pi a))^2 \dlot (2\pi-\kx)}
\label{eq:d4}
\end{equation}
The first of Eqs.(\ref{eq:37}) follows by keeping the first two terms in a series expansion of the right-hand side of Eq.(\ref{eq:d4}) in powers of $\dlo(2\pi-\kx)$. The distance $h_{n}$ in Eqs.(\ref{eq:37}) can then be derived as indicated by system (\ref{eq:31}).\\
\indent Eqs.(\ref{eq:16.1}) and (\ref{eq:53.1}) are obtained through similar treatments of the functions defined in Eqs.(\ref{eq:15}) and (\ref{eq:47}) respectively. In particular, for the bound states below the continuum, it appears possible
to give only the approximate value of the wavenumber $k^{+}$ while the wavenumber $k^{-}$ eludes the perturbation method due to a complicated dependence of its existence condition on the size of the cylinders. Similarly, in the case of two open channels, it turns out to be only possible to solve for the wavenumbers $k_{2n+1}$ whose existence is not subject to changes in cylinder sizes. Expressions
analogous to Eq.(\ref{eq:d4}) for the wavenumbers $k^{+}$ and $k_{2n+1}$ are,
\begin{subequations}\label{eq:d5}
\begin{equation}\label{eq:d5.1}
k^{+}\approx \sqrt{k^{2}_{x}-16\pi^2\dlot(\kx)}
\end{equation}
\begin{equation}\label{eq:d5.2}
k_{2n+1}\approx \sqrt{(2\pi+\kx)^2-16\pi^2\dlot(2\pi+\kx)}
\end{equation}
\end{subequations}
\indent To  establish Eq.(\ref{eq:56.1}), we recall that the point $k^{2n+1,l}_{x}$ is solution to the equation,
\begin{equation*}
\sqrt{\frac{k^{2}_{2n+1}-k^{2}_{x}}{k^{2}_{2n+1}-(2\pi-\kx)^2}}=\frac{l}{2n+1}
\label{eq:d6}
\end{equation*}
Hence,
\begin{equation*}
4\pi \kx (1-2r^2)+4\pi^2=(1-r^2)\Delta k, \quad r=\frac{l}{2n+1}
\label{eq:d7}
\end{equation*}
where $k^{2}_{2n+1}=(2\pi+\kx)^2-\Delta k$ for $\Delta k$ given by Eq.(\ref{eq:d5.2}). Thus,
\begin{equation*}
4\pi \kx(1-2r^2)+4\pi^2=\pi^2 u (1-r^2)(2\pi+\kx)^4, \quad u=R^4(\eec-1)^{2}
\label{eq:d8}
\end{equation*}
One can then look for a series solution $\kx=a_{0}+a_{1}u+a_{2}u^2+...$ This leads to Eq.(\ref{eq:56.1}).

\end{document}